\documentclass[11pt,a4paper]{article}
\usepackage{jheppub}
\usepackage{graphicx}

\def\be{\begin{equation}}
\def\ee{\end{equation}}
\def\bea{\begin{eqnarray}}
\def\eea{\end{eqnarray}}

\title{Probing Majorana neutrinos in rare $\pi^+ \to e^+ e^+ \mu^- \nu $ decays}
\author[a]{ Gorazd Cveti\v{c} }
\emailAdd{gorazd.cvetic@usm.cl}
\author[a]{, Claudio Dib }
\emailAdd{claudio.dib@usm.cl}
\author[b]{and C. S. Kim }
\emailAdd{cskim@yonsei.ac.kr} 
\affiliation[a]{Centro Cient\'{i}fico-Tecnol\'ogico de Valpara\'{\i}so
and  Department of Physics, \\
Universidad T\'ecnica Federico Santa Mar\'{\i}a, Valpara\'{\i}so, Chile}
\affiliation[b]{Dept. of Physics and IPAP, Yonsei University, Seoul 120-749, Korea }

\abstract{
We study the rare decays of charged $\pi$ mesons,
$\pi^+ \to e^+ e^+ \mu^- \bar\nu_\mu $ and $\pi^+ \to e^+ \mu^- e^+ \nu_e $
induced by a sterile neutrino $N$ with a mass in the range
$m_\mu < m_N < m_\pi$. The first process violates Lepton Number by two units and so occurs only if
$N$ is Majorana, while the second process conserves Lepton Number and occurs irrespective of the
Majorana or Dirac character of $N$.
We study a way to distinguish the Majorana vs. Dirac character
of $N$ in these processes using the muon spectrum.
We also find that the branching ratios could be at the reach of high luminosity
experiments like Project X at FNAL or any proposed neutrino (or muon) factories
worldwide.
}

\begin{document}

\hspace{9.cm}USM-TH-299, arXiv:1203.0573v2

\maketitle 

%
%
\section{Introduction}

One of the outstanding issues in neutrino physics today is to
clarify the  Dirac or Majorana character of neutrino masses.
The discovery of neutrino oscillations indicates that neutrinos
are massive particles with masses likely to be much smaller than
those of charged fermions \cite{SuperK1, SuperK2, SuperK3, SuperK4, SuperK5, SuperK6, SuperK7, SuperK8}.
This fact provides an important clue on the existence of a more
fundamental physics  underlying the Standard Model (SM) of
particle physics, because neutrinos are naturally massless in the
SM.

If neutrinos are Dirac particles, they must have right-handed
electroweak singlet components in addition to the known left-handed
modes; in such case Lepton Number remains as a conserved quantity.
Alternatively, if neutrinos are Majorana particles, then a
neutrino is indistinguishable from its antiparticle and Lepton
Number is violated by two units ($\Delta L=2$) in some
processes that involve neutrinos.
The experimental results to date are unable to distinguish between
these two alternatives.

There have been several attempts to determine the
Majorana nature of neutrinos by searching for $\Delta L=2$
processes. The most prominent of these
processes are neutrino-less double beta decays ($0\nu \beta
\beta $) in nuclei, which are regarded as the most sensitive way to
look for Lepton Number violation (LNV) \cite{Racah:1937qq,nndb1, nndb2, nndb3} 
(for a more recent review see \cite{Elliot}).
However, it has long been recognized that, even though the $0\nu \beta \beta $
experiments are very sensitive, the extraction of the neutrino
mass scale from nuclear $0\nu\beta \beta $ is a difficult task, because reliable information on
neutrino properties can be inferred only if the nuclear matrix
elements for $0\nu \beta \beta $ are also calculated reliably.
Even in the most refined treatments, the
estimates of the nuclear matrix elements remain affected by
various large uncertainties \cite{BB0n1, BB0n2, BB0n3, BB0n4, BB0n5, BB0n6}.

Another way to detect the Majorana nature of neutrinos
is to consider specific scattering processes with possible lepton number violation
in hadronic accelerators \cite{Keung:1983uu,Tello:2010am,Nemevsek:2011aa,Senjanovic:2011zz}.

Yet another avenue to detect the Majorana nature of neutrinos is to study $\Delta L=2$
rare meson decays \cite{Littenberg, Kova1, Ali, deltal2, Atre, cdkk}. 
%
%
%
%
%
%
%
In our previous work \cite{cdkk} we have studied rare 
$K$, $D$ and $B$ meson decays that are sensitive to neutrino masses above $m_\pi$, up to the masses of $B$ mesons.
The expected rates for the different decays are quite diverse, due to the kinematics of the processes and the current bounds on 
neutrino masses and mixing parameters. 
In this paper we want to focus our study on  the $\Delta L = 2$ decay of
charged $\pi$ mesons. This process is sensitive to a rather narrow neutrino mass range (between $m_\mu$ and $m_\pi$). However narrow, this mass range has good discovery potential in proposed high intensity
experiments,  such as Project X \cite{Project-X} of FNAL or at other neutrino factories or muon factories/colliders proposed worldwide. These kinds of experimental facilities will have very large samples of pion decays, much larger than the samples of any other meson decays, so they could be sensitive to much smaller decay rates, and so increase the potential for discovery of sterile neutrinos in this mass range. 
%
%
%
%
%
%
%
%

Specifically, here we consider the decay
$\pi^+ \to e^+ e^+ \mu^- \bar\nu_\mu $. The observation of this exclusive decay 
will clearly indicate Lepton Number violation. However,
because the final neutrino flavor cannot be observed in the experiment, 
a similar but Lepton Number conserving (LNC), $\Delta L =0$ (albeit 
Lepton Flavor violating) process,
$\pi^+ \to e^+ \mu^- e^+  \nu_e $, cannot be easily separated from the signal.
We explore how to distinguish these two decays in order to confirm the 
$\Delta L= 2$ process.
We must add that, by specifically choosing $\mu^-$ in the final state of 
the decay of a $\pi^+$ (i.e. a muon with opposite charge to that of the pion), 
one avoids a serious background coming from the radiative decays 
$\pi^+ \to \mu^+ \nu_\mu + \gamma^*(\to e^+ e^-)$.

The $\Delta L= 2$ pion decay we propose can only occur via a Majorana neutrino
in the intermediate state, and thus
its experimental observation can establish the Majorana
character of neutrinos and the absolute scale of their
masses in much the same way as in nuclear $0\nu \beta \beta $
decays, but there are some essential differences.
While the theoretical uncertainties in meson decays
are much easier to handle than in nuclear $0\nu \beta \beta$ decays, the latter are more realistic
options for experiments, as one can count with macroscopically
large samples of decaying nuclei. Indeed, $\Delta L=2$ meson decay rates, at least
in the case of light neutrinos ($m_\nu < 2$ eV), are prohibitively small for any
beam experiment where the mesons must be produced.

Nevertheless, in the case of a Majorana neutrino (denoted as $N$) with a mass in the intermediate
range $m_\mu < m_N < m_\pi$, a decay like
$\pi^+ \to e^+ e^+ \mu^- \bar\nu_\mu $ is dominated by a resonantly enhanced $s$-channel
amplitude \cite{Kova1, Kova, cdkk}: the intermediate neutrino $N$ goes on
its mass shell, making the decay rate large enough to be within reach of future experiments.

Moreover, since the intermediate neutrino $N$ is on its mass shell and $N$ is generally long lived, the process actually separates
in the sequence $\pi^+\to e^+ N$ followed by $N\to e^+\mu^-\nu$. The study of the first subprocess may already
allow to improve the current upper bounds on the magnitude of the lepton mixing element $|B_{e N}|$ for $m_N$ in the specified mass range.

Besides the effect on decay processes, there is much motivation to search for non-standard
neutrinos with masses of the order of ${\cal O} (10^0 - 10^2)$ MeV because their
existence has nontrivial observable consequences for cosmology and
astrophysics, in particular, they are presumed
to be a component of the Dark Matter in the Universe \cite{sdm1, sdm2}.
Actually, important bounds on neutrino masses and their
mixing with the standard neutrinos arise from Cosmology
and Astrophysics \cite{sb1}. There are also
laboratory bounds on sterile neutrinos coming from their contribution
to various processes that are forbidden in the SM, but those bounds turn out to
be much weaker than the cosmological and astrophysical bounds, except in specific cases
where the latter become inapplicable (see \cite{sb2} and references therein).

We should remark that neutrinos in our mass range of interest must be sterile
with respect to weak interactions.  From cosmological constraints we know that the standard
electron neutrino mass is below 2 eV \cite{cosmology}, while neutrino oscillation experiments
tell us that all three neutrino masses differ from one another by much less than that value
\cite{neutrino-osc} (for a recent review see \cite{neutrino-osc2}), and measurements of the $Z^0$ boson 
width constrain the number of standard neutrinos to three.
Consequently, all neutrinos with masses above 2 eV are assumed to be non-standard, and thus sterile with respect to
weak interactions.  In this work we will denote these neutrinos by the letter $N$, and reserve the letter $\nu$
for the standard neutrinos.

In section 2 we show the calculation of the decay rate of interest, 
$\pi^+ \to e^+ e^+ \mu^- \bar\nu_\mu $, and its related ``background" 
$\pi^+ \to e^+ e^+ \mu^-\nu_e $. We numerically explore the processes and discuss 
the results in section 3, in particular to study the ways to distinguish 
the Majorana vs. Dirac character of the intermediate neutrino using the 
muon spectrum. In section 4 we state our conclusions.

%
%

\section{Calculation of $\pi^+ \to e^+ e^+ \mu^- \bar\nu_\mu $ and $\pi^+ \to e^+ e^+ \mu^- \nu_e $}

\begin{figure}[htb] 
\centering\includegraphics[width=60mm]{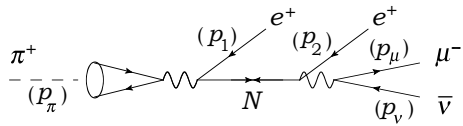}
\caption{\footnotesize Diagram for the Lepton Number violating process
$\pi^+\to e^+ e^+ \mu^-\bar\nu_\mu$ mediated by an on-shell Majorana neutrino $N$ of mass between $m_\mu$ and $m_\pi$.}
\label{FigPiMaj}
\end{figure}

\begin{figure}[htb] 
\centering\includegraphics[width=60mm]{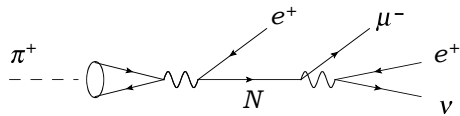}
\caption{\footnotesize Diagram for the Lepton Flavor violating 
(yet Lepton Number conserving) process
$\pi^+\to e^+ e^+ \mu^-\nu_e$ mediated by an on-shell neutrino $N$ of mass 
between $m_\mu$ and $m_\pi$.}
\label{FigPiDir}
\end{figure}

The decay $\pi^+ \to e^+ e^+ \mu^- \bar\nu_\mu $ is forbidden within the SM as it
violates lepton number by two units. However, if we assume the existence of
a Majorana neutrino $N$ with a mass in the range
$m_\mu < m_N < m_\pi$ and which mixes with the standard flavors, this decay
becomes dominated by the amplitude shown in figure \ref{FigPiMaj}, where the
intermediate Majorana neutrino $N$ goes on its mass shell, i.e., it is
produced and then decays.
Contributions where $N$ is not on its mass shell
are prohibitively small for any foreseeable experiment, so we will not
consider them here.
Since the final state neutrino flavor is not detected in these kinds of 
experiments, the process $\pi^+ \to e^+ e^+ \mu^- \nu_e$, shown in 
figure \ref{FigPiDir}, should be added in the rate. This latter process is 
also mediated by a neutrino $N$, but irrespective of its Majorana or Dirac 
character.
Consequently, if $N$ is a Majorana neutrino, both processes should occur,
while if it is a Dirac neutrino,  only the latter process is possible.

Given that the decay of the intermediate neutrino $N$ is necessarily very
weak, the two positrons in the process can actually be distinguished by
their spatially displaced vertices, so there is no need to consider the
diagram with crossed positron lines (this would not be the case if
$N$ were off-mass shell).
\footnote{If the crossed amplitude is included
for an on-shell $N$, the interference term turns
out to be zero and the two absolute squares give
the same contribution, so with the inclusion of the
symmetry factor $(1/2)$ one obtains the same result
as with a single diagram.}

\begin{figure}[htb] 
\centering\includegraphics[width=80mm]{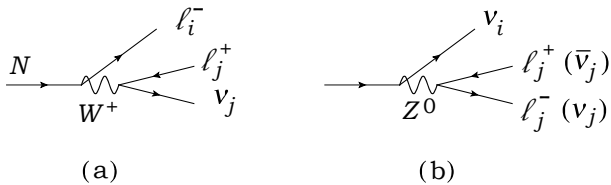}
\caption{\footnotesize Main decay of a massive neutrino $N$ with $m_N < m_{\pi}$:
(a) charged current channels, where the charged leptons can
be $e^+ e^-$, $e^+\mu^-$ or $\mu^+ e^-$, and (b) possible neutral
current channels, where the charged lepton pair can only
be $e^+ e^-$.}
\label{NdecW}
\end{figure}

We can estimate the expected separation of the vertices by calculating the
lifetime of the neutrino $N$. If $N$ is a Dirac neutrino,
its main decay modes are given in figure \ref{NdecW}, mediated by (a) charged and (b) neutral weak currents. There are additional (smaller) contributions from radiative channels which we neglect.

The dominant charged current mode is  $N\to e^+e^-\nu$,
while the channels with a muon in the final state are suppressed by
an extra factor $f(m_\mu^2/m_N^2)$, where $f(x)$ is the well-known 3-body
function that also appears, for example, in muon decay
\begin{equation}
f(x) = 1-8x+8x^3-x^4-12x^2\ln x \ .
\label{Ffunction}
\end{equation}
This factor, when $m_N$ increases from $m_{\mu}$ to $m_{\pi}$,
decreases from $7.3 \times 10^{-3}$ to zero, so we can neglect the channels with a final muon. Concerning the neutral current channels, these appear only in some models of neutrino mixing, and include mixing elements other than the $B_{\ell N}$. In models where they appear, they are at most comparable in size to the charged ones. We will then consider these neutral channels only implicitly by a factor ${\cal K}$, as shown below.

Now, if $N$ is a Majorana neutrino, the channels
charge-conjugate to those of figure \ref{NdecW}.a contribute as well,
and therefore the witdh $\Gamma_N$ is twice as large as
in the case of Dirac neutrino.
Therefore, the decay width of the neutrino $N$ in our cases of interest can be expressed in general as:
\begin{equation}
\Gamma_N \approx \Gamma(N \to e^+e^-\nu) \ {\cal C} \ {\cal K}\ \approx
|B_{eN}|^2 \frac{G_F^2 m_N^5}{192\pi^3}\  {\cal C}\  {\cal K} \ ,
\label{DNwidth}
\end{equation}
where ${\cal C} =1$ or $2$ if $N$ is a Dirac or Majorana neutrino, respectively, and ${\cal K}$
represents a correction to include the
neutral current interaction channels shown in figure \ref{NdecW}.b (${\cal K}$ would be unity  
if the neutral channels were absent). 
%
%
%
%
%
%
Using the results of ref.~\cite{Atre}, 
Appendix C, ${\cal K}$ turns out to be
\begin{equation}
{\cal K} = 1.757 + 0.606 \frac{\left( |B_{\mu N}|^2
+ |B_{\tau N}|^2 \right)}{|B_{e N}|^2} \ ,
\label{calK}
\end{equation}
where $\sin^2 \theta_W = 0.231$ was used.
The decays contained in this expression are $N \to \nu_i e^+ e^-$,
$N \to {\bar \nu}_i e^+ e^-$ and $N \to \nu_i \nu_j {\bar \nu_j}$,
where $i,j =1, 2, 3$ label the light mass eigenstate neutrinos.
%
%
%
%
%
We denote by $B_{\ell N}$ the mixing coefficient
between the standard flavor neutrino $\nu_{\ell}$ ($\ell = e, \mu, \tau$)
and the (heavy) mass eigenstate $N$:
\begin{equation}
\nu_{\ell} = \sum_{j=1}^3 B_{\ell \nu_j} \nu_j + B_{\ell N} N \ .
\label{mixing}
\end{equation}
Only one heavy neutrino is included, for simplicity of notation.
The extension of our results to scenarios with several heavy neutrinos
is usually straightforward.

Numerically, the decay distance $c \tau_N$ for a such a Majorana neutrino ranges from about $300/|B_{eN}|^2$ meters for $m_N =m_\pi$ to about 4 times longer for $m_N =m_\mu$, without considering relativistic dilations. Taking into account the current upper bounds $|B_{eN}|^2\lesssim 10^{-8}$ \cite{PIENU:2011aa, Atre}, decay lengths above $10^9$ meters are expected, so in most cases the vertex separation is enormous. For a detector of length $L$ (with $L \ll c \tau_N$), the probability of a neutrino $N$ to decay inside the detector is $P_N = L / \gamma c \tau_N$ (e.g. $P_N\sim 10^{-7}/\gamma$ for a detector about 100 m long), where $\gamma$ is the relativistic dilation factor.

Clearly the search for a massive neutrino $N$ in these processes should first be done by looking at the energy spectrum of the positrons produced in the primary vertex, i.e. in a search for the pion decay mode $\pi^+ \to e^+ N$. For $m_N$ in the range $(m_\mu, m_\pi)$, the energy of these positrons (in the pion rest frame) is rather low: $E_e  = (m_\pi^2-m_N^2)/2m_\pi$, which is at most $30$ MeV, for $m_N=m_\mu$. Notice that, because $m_N$ is not much less than $m_\pi$, this decay is not chirally suppressed (unlike $\pi^+ \to e^+ \nu_e$). Using the definition of the pion decay constant $f_\pi$ in terms of the charged axial current, $\langle 0 |\bar u \gamma^\mu (1-\gamma_5)d|\pi^+(p)\rangle = i f_\pi p^\mu $, and the usual notation for the mixing of the massive neutrino $N$ with a lepton flavor $\ell$ by $B_{\ell N}$, the decay rate of this mode is:
\begin{equation}
\Gamma(\pi^+\to e^+ N) = \frac{1}{8\pi} G_F^2 f_\pi^2 |V_{ud}|^2 \frac{m_N^2(m_\pi^2-m_N^2)^2}
{m_\pi^3}|B_{eN}|^2,
\label{piondecaymode}
\end{equation}
which gives a branching fraction (approximating the total width of the pion by the decay rate of its dominant channel $\pi^+\to\mu^+\nu_\mu$):
\begin{equation}
Br(\pi^+\to e^+N) \simeq \frac{m_N^2(m_\pi^2-m_N^2)^2}{m_\mu^2(m_\pi^2-m_\mu^2)^2}|B_{eN}|^2.
\label{piontoeN}
\end{equation}
Consequently, the suppression of this low energy positron peak comes almost entirely from the mixing element $|B_{eN}|^2$, except for an additional phase space suppression $(1-m_N^2/m_\pi^2)^2$ if $m_N$ is near the pion mass threshold. There is no chiral suppression here, unlike in $\pi^+\to e^+\nu_e$, because $N$ is heavy.

As a preliminary experimental analysis, the study of the positron energy spectrum in the primary decay $\pi^+ \to e^+ N$, specifically the search for bumps in the spectrum below the peak coming from the standard process $\pi^+ \to e^+ \nu_e$, will allow the discovery of the heavy neutrino $N$, or otherwise to put more stringent upper bounds on the lepton mixing element $|B_{eN}|$.

Although the observation of the positron in the primary decay 
$\pi^+\to e^+ N$ can lead to the discovery of the heavy neutrino, it cannot 
tell about its Majorana or Dirac character. For this latter purpose, 
we must observe the subsequent $N$ decay. Let us then consider the 
Lepton Number violating decay of interest, $\pi^+\to e^+ e^+\mu^- {\bar \nu}_{\mu}$, 
depicted in figure \ref{FigPiMaj}.  
%
%
%
%
%
%
%
%
In what follows we neglect the electron and standard neutrino masses.

Since the process is dominated by the intermediate neutrino $N$ on mass shell, the transition probability can be estimated using the 
narrow width approximation. The details of the derivation of the rates are given in Appendix.  
After integrating the transition rate over all final particles momenta
but the muon energy, we obtain the spectrum of the muon energy $E_\mu$ [see eq.~(\ref{spectrumLNV})]:
\begin{eqnarray}
\lefteqn{\Gamma(\pi^+\to e^+e^+\mu^-\bar\nu_\mu) = \int_{m_\mu}^{ \frac{m_N^2 + m_\mu ^2 }{2 m_N}  }
  d E_{\mu} }
\label{spectrum1LNV}\\
 && \qquad \frac{ G_F^4  |B_{eN}|^4 f_\pi^2 \, |V_{ud}|^2 \, m_N^2\,  (m_\pi^2- m_N^2)^2 }{16\pi^4  \,  m_\pi^3 \,  \Gamma_N}
  \  E_{\mu}\
  (m_N^2 + m_\mu^2- 2 m_N  E_{\mu})
  \sqrt{E_{\mu}^2 - m_\mu^2},
\nonumber
\end{eqnarray}
where $\Gamma_N$ is given in Eq.\ (\ref{DNwidth}), with ${\cal C}=2$ as here $N$ is a Majorana neutrino.
As a reference, the general differential decay rate
$d \Gamma/d E_{\ell}$ for a general pseudoscalar decay of this type,
$M^+ \to \ell_1^+ \ell_2^+ \ell^- {\bar \nu}_{\ell}$, with general nonzero masses
of the particles, is given in eq.~(\ref{dGMEel}) in the Appendix.

The appearance of $\Gamma_N$ in the denominator of these expressions is 
due to the dominance of $N$ on its mass shell.
Indeed, as in all narrow width approximations, the rate is proportional 
to the branching ratio of the $N$ decay subprocess, namely:
\[
\Gamma(\pi^+\to e^+ e^+ \mu^-\bar\nu_\mu) =  \Gamma(\pi^+\to e^+ N)\cdot Br(N\to e^+ \mu^- \bar\nu_\mu),
\]
where $\Gamma(\pi^+\to e^+ N)$ is given in eq.\ (\ref{piondecaymode}), $Br(N\to e^+\mu^-\nu_\mu) \equiv \Gamma(N\to e^+\mu^-\nu_\mu)/\Gamma_N$,  and  $\Gamma(N\to e^+ \mu^-\bar\nu_\mu)$ is just like the expression for $\Gamma(N\to e^+ e^-\nu)$ given in eq.\ (\ref{DNwidth}) times a factor $f(m_\mu^2/m_N^2)$ coming from
the phase space with a muon instead of an electron. Given the total width $\Gamma_N$ in eq.\ (\ref{DNwidth}), the resulting branching ratio is:
\[
Br(N\to e^+\mu^-\bar\nu) = \frac{1}{{\cal C K}} f(m_\mu^2/m_N^2),
\]
This branching ratio, for $m_N$ in the range $(m_\mu , m_\pi)$ (and ${\cal C} =2$ for a Majorana $N$) has a maximum value  $\sim 4\times 10^{-3}$ for $m_N = m_\pi$,
dropping quickly several orders of magnitude for lower masses, and vanishing at $m_N = m_\mu$. On the other hand, the \emph{production} of $N$ in the pion decay is phase-space suppressed as $m_N\to m_\pi$.

Now, $\pi^+\to e^+ e^+ \mu^-\bar\nu_\mu$ occurs only if $N$ is a Majorana, not 
Dirac, neutrino. However, as previously mentioned, since the final 
neutrino flavor is not experimentally detectable, there is a ``background'' 
for this process that is mediated by a neutrino $N$ regardless of 
its Majorana or Dirac character: $\pi^+ \to e^+ e^+ \mu^- \nu_e $. 
The diagram for this process is shown in figure \ref{FigPiDir}.  
Following a similar procedure as before, for this process we find 
the muon energy  spectrum 
in the $N$ rest frame [see eq.~(\ref{spectrumLFV})] to be:
\begin{eqnarray}
\lefteqn{\Gamma (\pi^+\to e^+\mu^- e^+\nu_e)
= \int_{m_\mu}^{\frac{m_N^2+m_\mu^2}{2 m_N}} dE_{\mu}
}
\label{spectrum1LFV}\\
&&\frac{ G_F^4 |B_{eN}B_{\mu N}^\ast|^2 f_\pi^2\, |V_{ud}|^2 \, 
m_N^2(m_\pi^2-m_N^2)^2}
{32\pi^4  \, m_\pi^3  \, \Gamma_N}
 \left\{(m_N^2+m_\mu^2)E_{\mu}
-\frac{2}{3}m_N(2E_{\mu}^2+m_{\mu}^2)\right\}\sqrt{E_{\mu}^2-m_{\mu}^2}.
\nonumber
\end{eqnarray}
%

Consequently, when the intermediate neutrino $N$ is Majorana, the
measurable muon energy spectrum is represented by the sum of expressions (\ref{spectrum1LNV}) and
(\ref{spectrum1LFV}):

\bea
\lefteqn{
\frac{d \Gamma^{(M)}}{d E_{\mu}} (\pi^+ \to e^+ e^+ \mu^- \nu)=
\frac{ G_F^4 |B_{eN}|^2 f_\pi^2\, |V_{ud}|^2 \,
m_N^2(m_\pi^2-m_N^2)^2}{16\pi^4  \, m_\pi^3  \, \Gamma_N}
\, \sqrt{E_{\mu}^2 - M_{\mu}^2}
}
\label{diffdecay}\\
&& 
\left\{ |B_{e N}|^2 E_{\mu} (m_N^2 + m_{\mu}^2 - 2 m_N E_{\mu})
+ |B_{\mu N}|^2 \left[ \frac{1}{2} E_{\mu} (m_N^2 + m_{\mu}^2)
- \frac{2}{3} m_N E_{\mu}^2 - \frac{1}{3} m_N m_{\mu}^2 \right]
\right\} ,
\nonumber
\eea
where $E_{\mu}$ is the muon energy in the neutrino $N$ rest frame,
and varies between $m_{\mu}$ and $(m_N^2+m_{\mu}^2)/(2 m_N)$.
For the decay width $\Gamma_N$, we must use eq.~(\ref{DNwidth}), with ${\cal C}=2$ because here $N$ is Majorana.

Notice that in this spectrum the differential rate is defined in the frame of the initial particle, i.e. the pion, while 
the variable $E_\mu$ is the muon energy in the neutrino $N$ rest frame. 
If one wanted to express $d\Gamma$ also in the $N$ rest frame, we should include the relativistic time dilation factor
$1/\gamma = 2 m_\pi m_N/(m_\pi^2 + m_N^2)$ for the moving pion in the $N$ frame, a factor which is between $0.9625$ and unity for $m_N$ between $m_{\mu}$ and $m_{\pi}$.  To avoid possible confusions, we normalize the spectrum relative to the pion decay width, thus obtaining a spectral branching ratio:
\begin{eqnarray}
\lefteqn{ \frac{d Br^{(M)}}{d E_{\mu}} (\pi^+ \to e^+ e^+ \mu^- \nu) \equiv
\frac{1}{\Gamma_{\pi^+}}\frac{d \Gamma^{(M)}}{d E_{\mu}} (\pi^+ \to e^+ e^+ \mu^- \nu)
}
\nonumber\\
&&=
\frac{1}{2\cal K} \frac{ (m_\pi^2-m_N^2)^2}{(m_\pi^2-m_\mu^2)^2} \frac{96}{m_N^3 m_\mu^2}
\,
\sqrt{E_{\mu}^2 - M_{\mu}^2}
\label{diffBr}\\
&& 
\left\{ |B_{e N}|^2 E_{\mu} (m_N^2 + m_{\mu}^2 - 2 m_N E_{\mu})
+ |B_{\mu N}|^2 \left[ \frac{1}{2} E_{\mu} (m_N^2 + m_{\mu}^2)
- \frac{2}{3} m_N E_{\mu}^2 - \frac{1}{3} m_N m_{\mu}^2 \right]
\right\} ,\nonumber
\end{eqnarray}
where we have approximated the total width of the charged pion, $\Gamma_{\pi^+}$, by its dominant (by far) decay channel, $\Gamma(\pi^+\to \mu^+ \nu_\mu)$.
Integrating this spectrum over the muon energy we obtain the corresponding branching ratio:
\begin{equation}
Br^{(M)}(\pi^+ \to e^+ e^+ \mu^- \nu)  =
  \frac{ |B_{e N}|^2 +  |B_{\mu N}|^2}{2\cal K} \  \frac{m_N^2 (m_\pi^2- m_N^2)^2 }{ m_\mu^2 (m_\pi^2- m_\mu^2)^2 }   f \left(m_\mu^2/m_N^2\right) \ .
\label{totalBr}
 \end{equation}

In contrast, if $N$ were a Dirac neutrino, the measured process would correspond solely to the Lepton Number conserving decay $\pi^+ \to e^+ e^+ \mu^- \nu_e$, and the muon spectrum would be given by eq.~(\ref{spectrum1LFV}), with $\Gamma_N$ given by eq.~(\ref{DNwidth}) with ${\cal C}=1$, appropriate for a Dirac $N$. The corresponding spectral and total branching ratios would then be given by:
\bea
 \frac{d Br^{(D)}}{d E_{\mu}} (\pi^+ \to e^+ e^+ \mu^- \nu_e)
&=&
\frac{|B_{\mu N}|^2}{\cal K} \  \frac{ (m_\pi^2-m_N^2)^2}{(m_\pi^2-m_\mu^2)^2} \frac{96}{m_N^3 m_\mu^2}
\label{diffBrD}
\\
&& \times \sqrt{E_{\mu}^2 - M_{\mu}^2}
\left\{  \frac{1}{2} E_{\mu} (m_N^2 + m_{\mu}^2)
- \frac{1}{3} m_N (2 E_{\mu}^2 + m_{\mu}^2)
\right\}  \nonumber
\eea
and
\begin{equation}
Br^{(D)}(\pi^+ \to e^+ e^+ \mu^- \nu_e)  =
  \frac{ |B_{\mu N}|^2}{\cal K} \  \frac{m_N^2 (m_\pi^2- m_N^2)^2 }{ m_\mu^2 (m_\pi^2- m_\mu^2)^2 }   f \left(m_\mu^2/m_N^2\right) \ ,
\label{totalBrD}
 \end{equation}
respectively.
%
%
%
%
%
%

%
%

\section{Numerical studies of branching ratios and spectra}
\label{num}

Here we want to analyze numerically the spectra and branching ratios for the processes $\pi^+ \to e^+ e^+ \mu^- \nu$
that we calculated in the previous section,  to see whether they can be accessed by experiment
within the specified range for $m_N$ and the possible values of the mixing elements $B_{\ell N}$.
As shown in the previous section, we express spectra and rates as
branching ratios, relative to the charged pion decay width.

In order to study the behavior of these quantities as a function of the neutrino mass, $m_N$,
it is convenient to factor out the dependence on the mixing elements. Here we do this
by defining the \emph{reduced} differential branching ratios as:
\be
d\overline{Br}  (\pi^+ \to e^+ e^+ \mu^- \nu) \equiv \frac{\cal K}{(|B_{e N}|^2 + |B_{\mu N}|^2)}
{d Br}  (\pi^+ \to e^+ e^+ \mu^- \nu),
\label{dcalG}
\ee
and similarly for the integrated branching ratios as well. We recall that  ${\cal K}$ ($\sim 1$) is the
correction factor defined in eq.~(\ref{DNwidth}),  which represents possible
neutral current channels in the massive neutrino decay width $\Gamma_N$.

With this factor extracted, the order-of-magnitude size of
our reduced branching ratios is given mainly by the mass of the 
intermediate neutrino $m_N$,
and not so much by the mixing elements $B_{\ell N}$:
the reduced branching ratios thus obtained from
eqs.\ (\ref{diffBr} --\ref{totalBrD}) depend on the mixing elements at most in the combinations
\be
 \alpha_M \equiv \frac{|B_{e N}|^2}{|B_{e N}|^2+|B_{\mu N}|^2} \quad \textrm{and}\quad
  1- \alpha_M \equiv \frac{|B_{\mu N}|^2}{|B_{eN}|^2+ |B_{\mu N}|^2} ,
\label{kD}
\ee
quantities which are bounded between 0 and 1.
\begin{figure}[b] 
\centering\includegraphics[width=100mm]{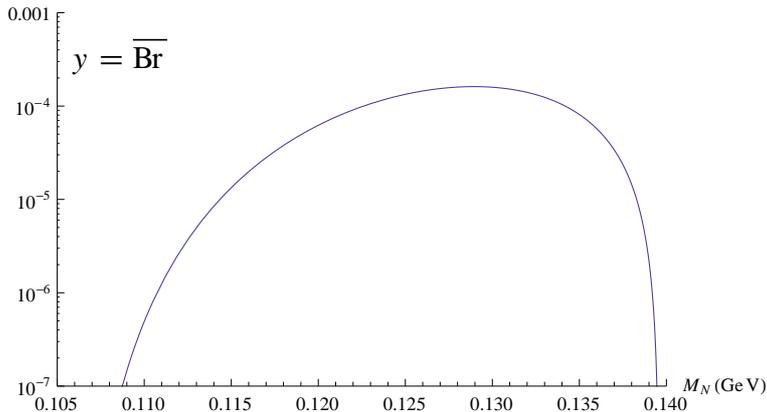}
\caption{\footnotesize The reduced branching ratio $\overline Br^{(M)}(\pi^+\to e^+e^+\mu^-\nu)$,
eqs.~(\ref{totalBr}) and (\ref{dcalG}),  as a function of the neutrino mass $m_N$.}
\label{plGMvsMN}
\end{figure}
In Table \ref{tab1} and figure \ref{plGMvsMN} we present numerical values for the reduced
branching ratio $\overline Br^{(M)}(\pi^+ \to e^+ e^+ \mu^- \nu)$, related to eq.\ (\ref{totalBr}), for various values of the neutrino mass $m_N$.
The true branching ratios mediated by a Majorana neutrino $N$  are obtained from the reduced ones after multiplying them by the factor
$(|B_{eN}|^2+ |B_{\mu N}|^2)/{\cal K}$, while if $N$ is of Dirac type the required factor is
$2 |B_{\mu N}|^2/{\cal K}$.

\begin{table}
\caption{Reduced branching ratio for the indicated process, induced by a 
Majorana neutrino of mass $m_N$ [\emph{cf.} eqs.\ (\ref{totalBr}) and (\ref{dcalG})].
}
\begin{center}
\label{tab1}
\begin{tabular}{|c|c|}
\hline
 $m_N$ [GeV] &
${\overline Br}^{(M)}(\pi^+ \to e^+ e^+ \mu^- \nu)$
\\
\hline
0.112 & $2.67 \cdot 10^{-6}$
\\
0.119 & $4.94 \cdot 10^{-5}$
\\
0.126 & $1.45 \cdot 10^{-4}$
\\
0.129 & $1.61 \cdot 10^{-4}$
\\
0.133 & $1.24 \cdot 10^{-4}$
\\
\hline
\end{tabular}
\end{center}
\end{table}

From the above we see that the measured branching ratio $Br(\pi^+ \to e^+ e^+ \mu^- \nu)$ is predicted to give
in general similar values whether it is mediated by a Majorana or a Dirac neutrino.
Consequently, the experimental measurement of this branching ratio
cannot represent a method for distinguishing between the Majorana and the Dirac character of
the intermediate neutrino, if this one exists in the relevant mass range.
On the other hand, the measurement of
the muon energy spectrum, i.e.
eqs.~(\ref{diffBr}) and (\ref{diffBrD}), may allow us to
distinguish between the two cases.

\begin{figure}[htb]
\begin{minipage}[b]{.49\linewidth}
\centering\includegraphics[width=\linewidth]{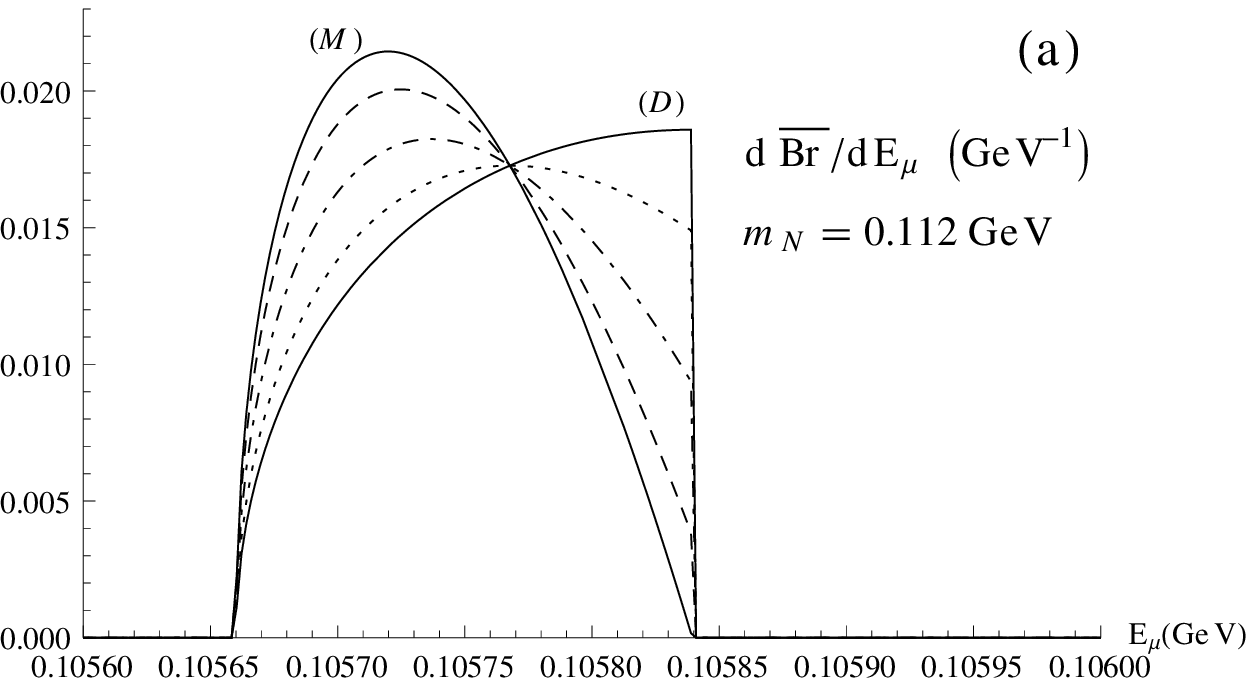}
\end{minipage}
\begin{minipage}[b]{.49\linewidth}
\centering\includegraphics[width=\linewidth]{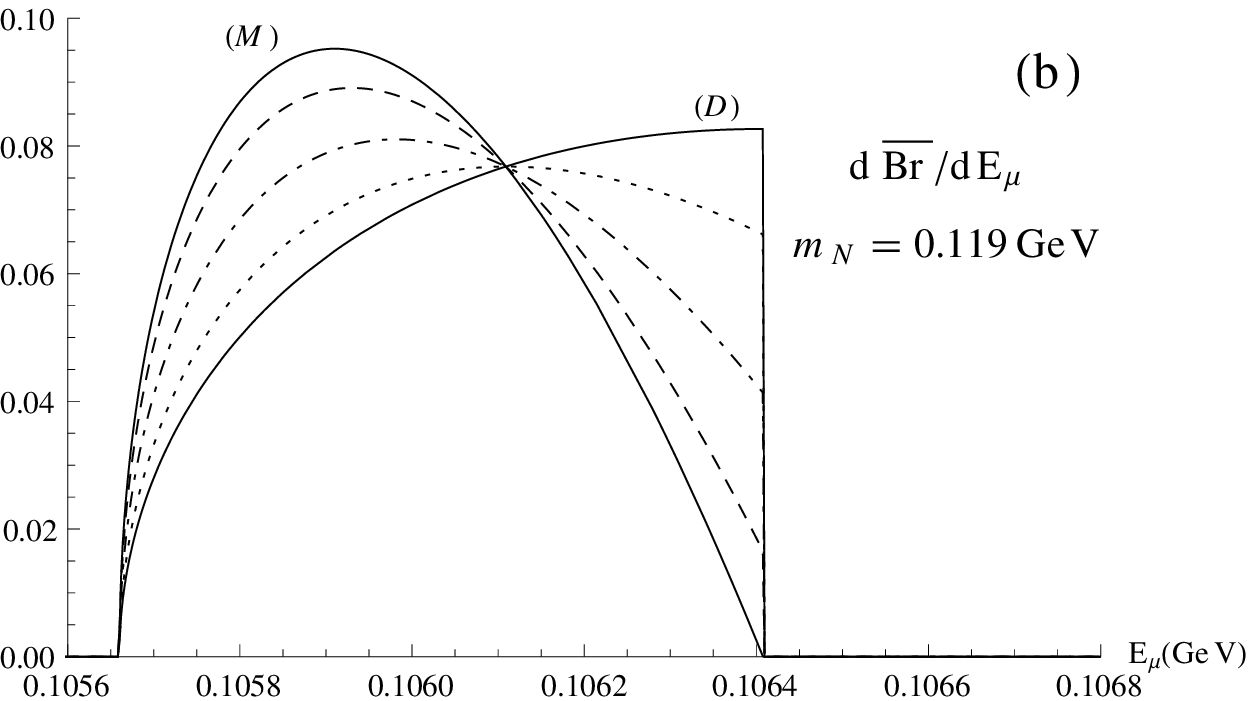}
\end{minipage}
\begin{minipage}[b]{.49\linewidth}
\centering\includegraphics[width=\linewidth]{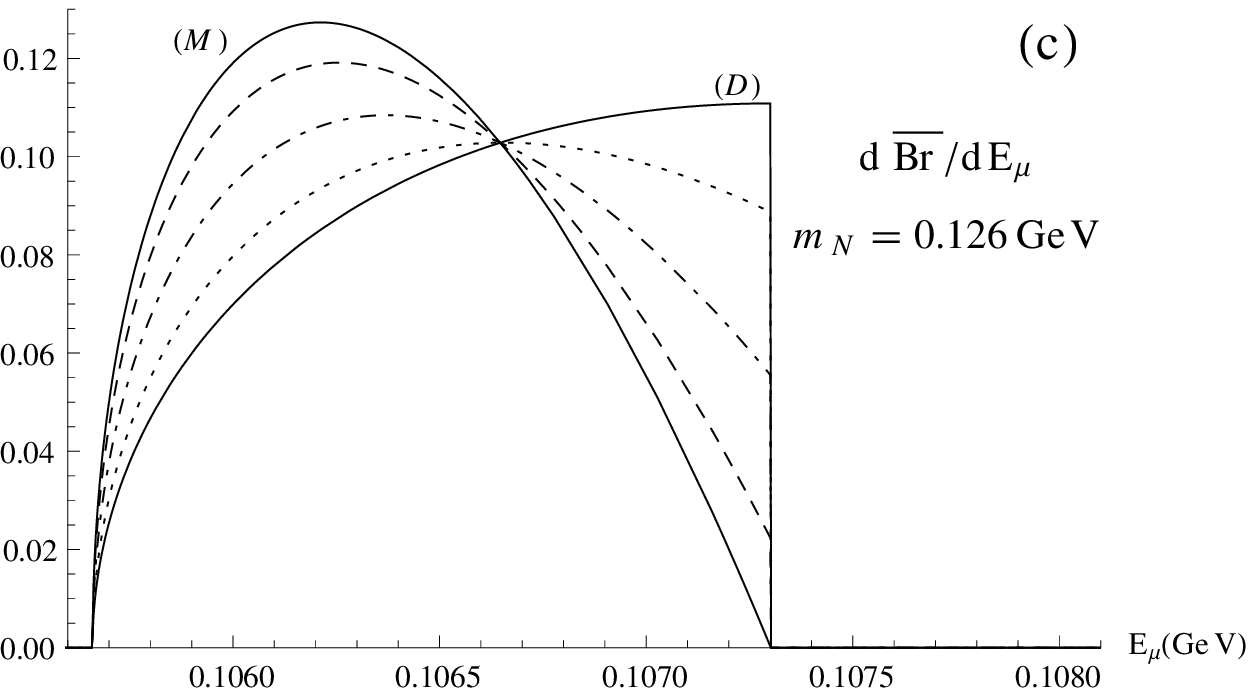}
\end{minipage}
\begin{minipage}[b]{.49\linewidth}
\centering\includegraphics[width=\linewidth]{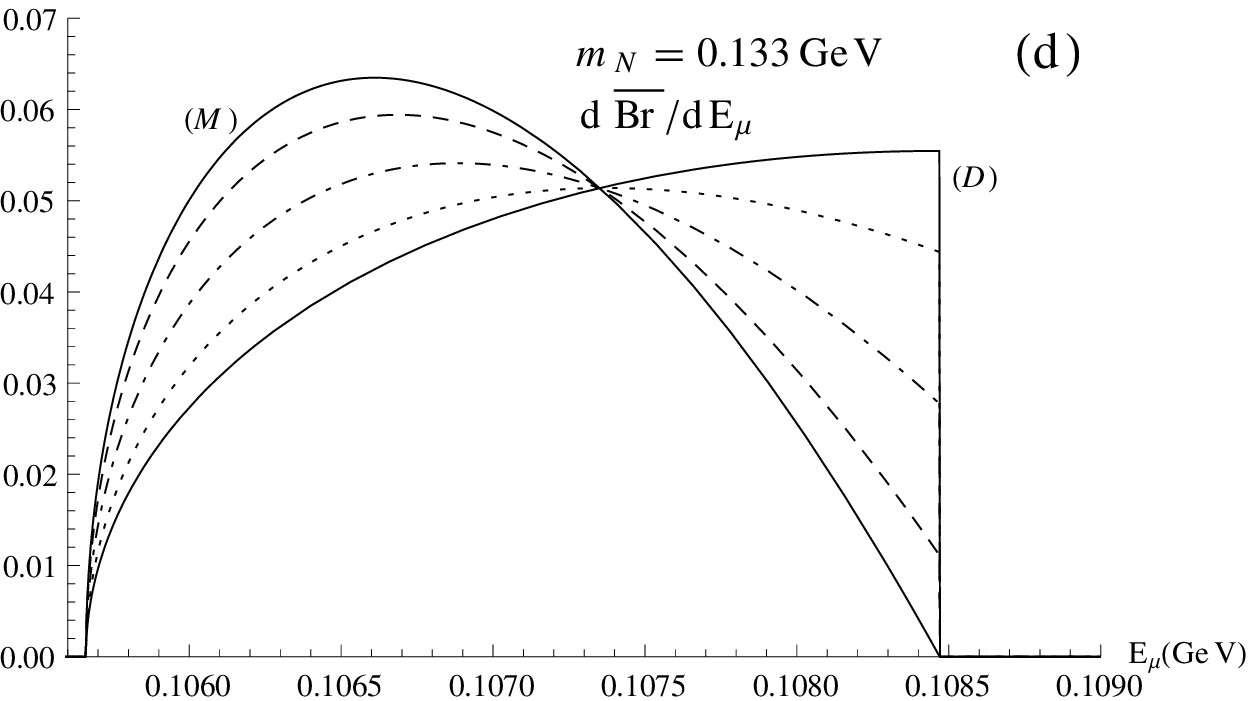}
\end{minipage}\vspace{-0.4cm}
\caption{\footnotesize The reduced differential branching ratio
$d \overline{Br}/d E_{\mu}$ as a function of the muon energy in the neutrino $N$ rest frame, $E_{\mu}$, as defined via
eqs.~(\ref{diffBr})-(\ref{dcalG}), for the decays
$\pi^+ \to e^+ e^+ \mu^{-} {\nu}$ mediated by a Majorana neutrino $N$, for various neutrino
masses: (a) $m_N=0.112$ GeV; (b) $m_N=0.119$ GeV; (c) $m_N=0.126$ GeV; (d) $m_N=0.133$ GeV.
In each graph there are five curves, corresponding to different values of the admixture parameter $\alpha_M$ [eq.~(\ref{kD})]:
$\alpha_M=1.0$ is the solid (M) curve; $0.8$ (dashed); $0.5$ (dot-dashed); $0.2$ (dotted). The case mediated by a Dirac neutrino
is also presented as the solid line labelled (D), with the distribution normalized so that the area under the curve is the same as in
the Majorana cases.}
\label{dGdEelCMN}
\end{figure}
\begin{figure}[htb]
\begin{minipage}[b]{.49\linewidth}
\centering\includegraphics[width=\linewidth]{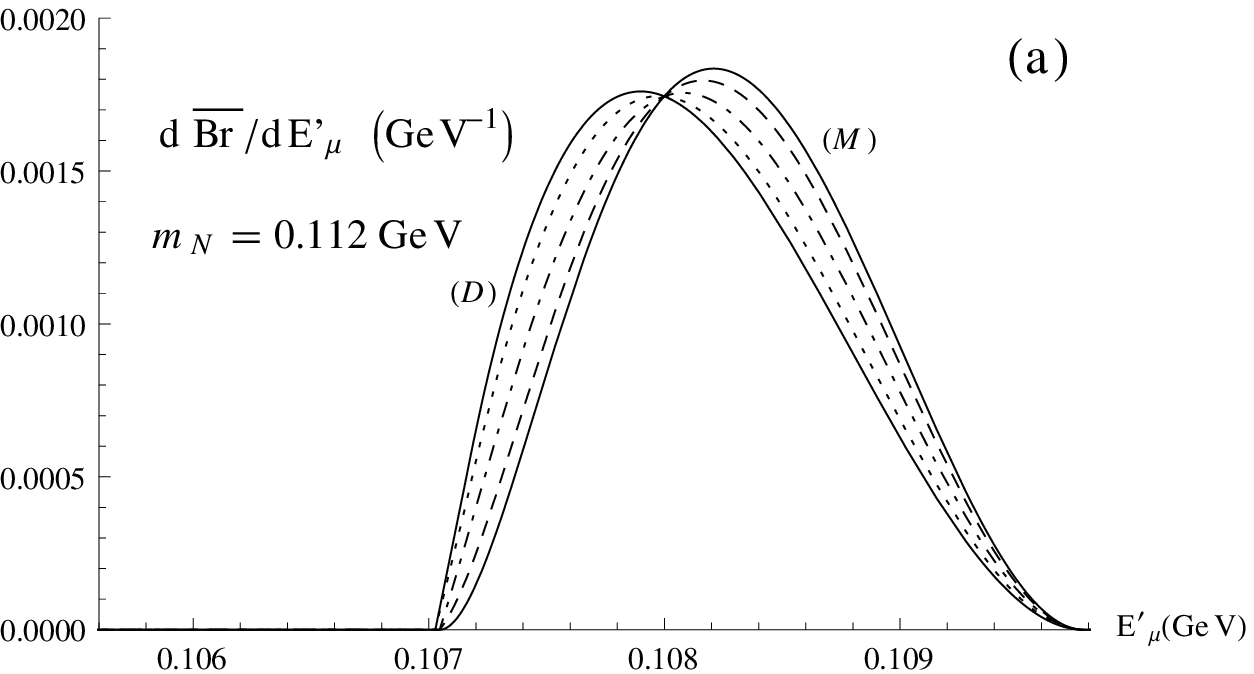}
\end{minipage}
\begin{minipage}[b]{.49\linewidth}
\centering\includegraphics[width=\linewidth]{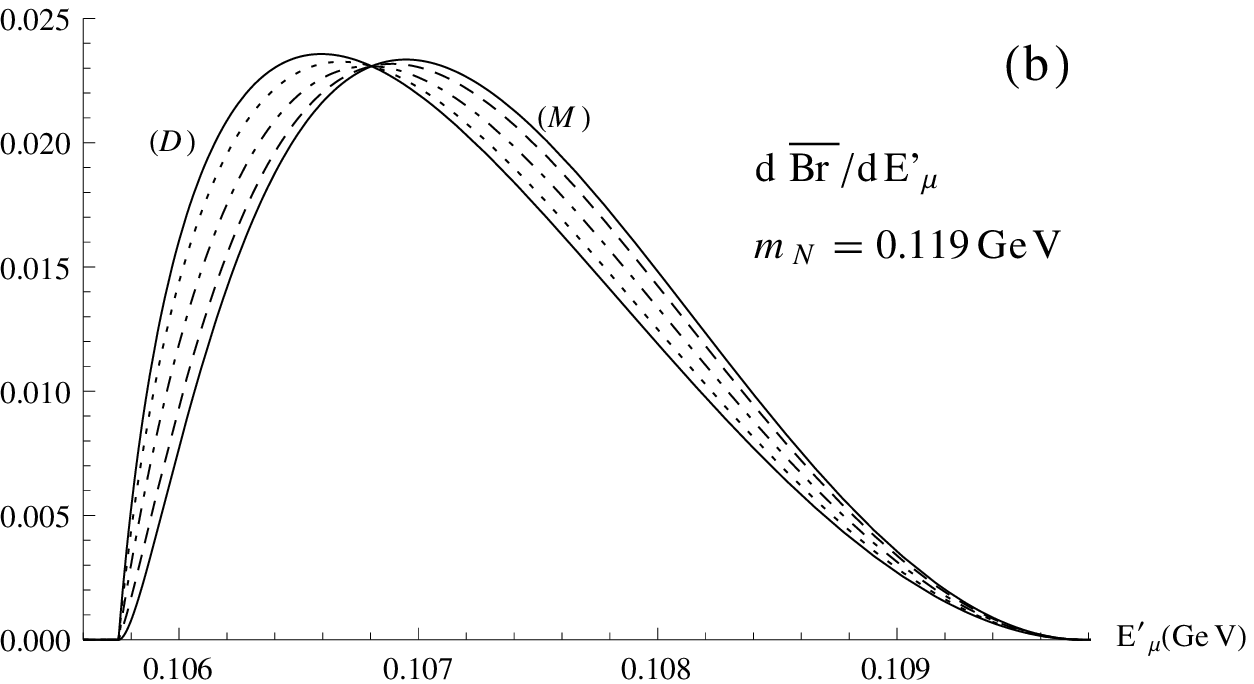}
\end{minipage}
\begin{minipage}[b]{.49\linewidth}
\centering\includegraphics[width=\linewidth]{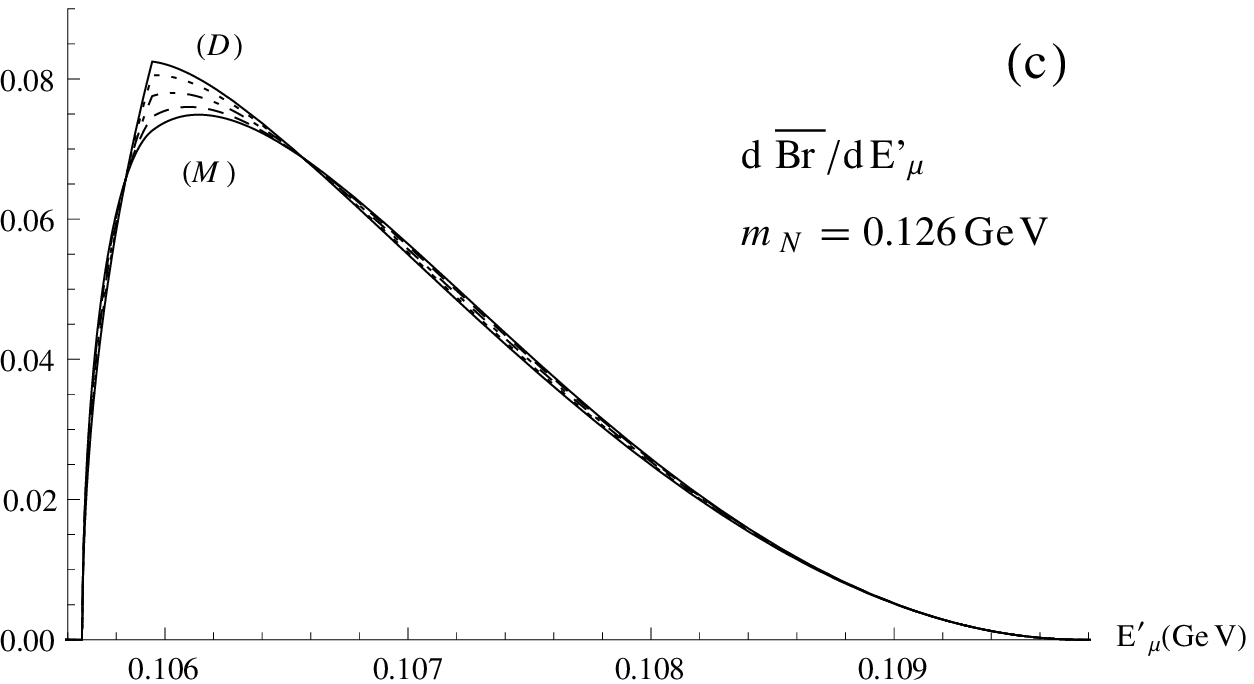}
\end{minipage}
\begin{minipage}[b]{.49\linewidth}
\centering\includegraphics[width=\linewidth]{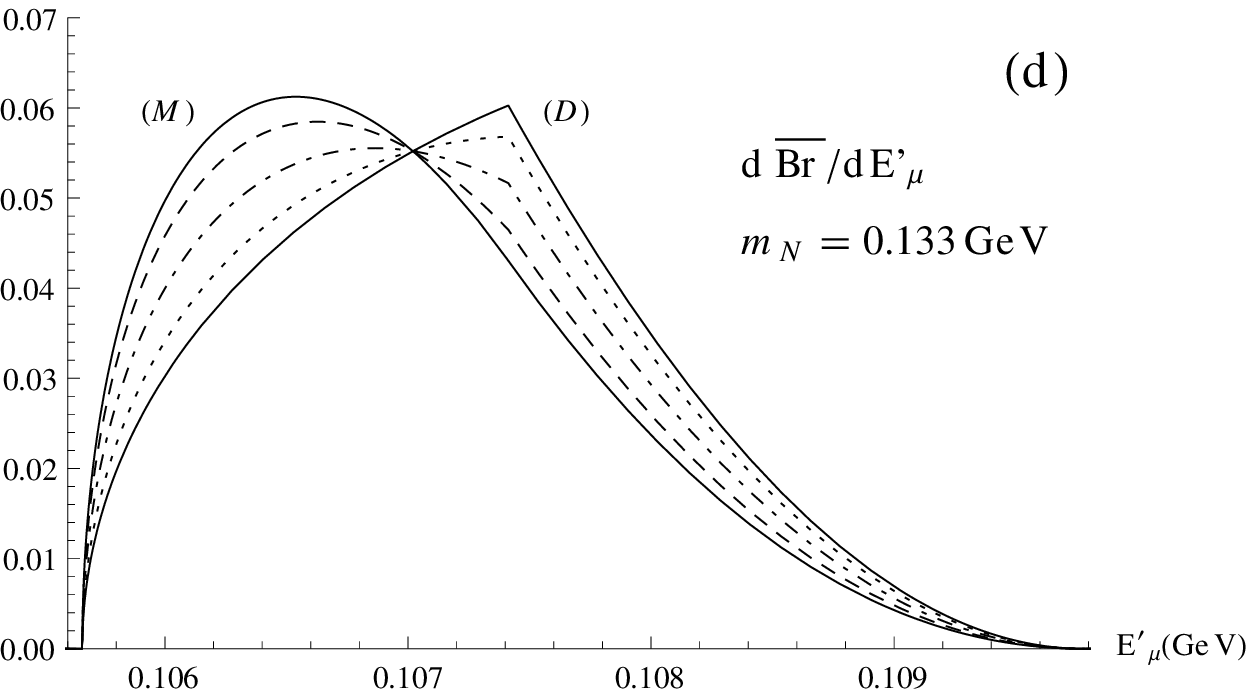}
\end{minipage}\vspace{-0.4cm}<
\caption{\footnotesize Same spectral distributions as in figures \ref{dGdEelCMN}, but now
for the muon energy $E_{\mu}^\prime$ measured in the rest frame of the pion.}
\label{dGdEelCMM}
\end{figure}

In figures \ref{dGdEelCMN} we present the reduced spectra ${d\overline Br^{(M)}}/d E_{\mu}$ as
a function of the muon energy in the intermediate neutrino $N$ rest frame. In each graph, four of the five curves correspond to different values of the
``admixture parameter'' ($\alpha_M = 1,$ 0.8, 0.5 and 0.2) in the Majorana $N$ case, and
the fifth curve correspond to the Dirac $N$ case, ${d\overline Br^{(D)}}/d E_{\mu}$.
Each graph corresponds to a given representative value of the intermediate neutrino
mass ($m_N =0.112$, 0.119, 0.126 and 0.133 GeV
in figures \ref{dGdEelCMN}(a)-(d), respectively).
Note that $m_N$ must be between
$m_{\mu} \approx 0.105$ GeV and $m_{\pi} \approx 0.140$ GeV for this process to occur.
 Consequently the kinetic energy of the muon in this frame runs over a very limited range, namely from zero up to $(m_N-m_\mu)^2/2m_N$.  This upper end is at most $\sim 4.3$ MeV if $m_N \to m_\pi$ and vanishes as $m_N \to m_\mu$.

Concerning these spectra, as already mentioned, $E_{\mu}$ is the muon energy in the rest frame of the intermediate neutrino $N$.
One may consider more practical to use the muon energy in the pion rest frame,
which here we denote as  $E^{\prime}_{\mu}$
(in the Appendix we treat in detail the transformation
$d \Gamma/d E_{\mu} \mapsto d \Gamma/d E^{\prime}_{\mu}$).
In figures \ref{dGdEelCMM}(a)-(d), we show the spectra of the same cases, respectively, but as functions
of the muon energy in the pion rest frame, ${d\overline{Br}}/d E^\prime_{\mu}$.

The $\alpha_M =1$ case is denoted in these figures by the solid line with the label (M), because for this admixture the Lepton Number conserving process $\pi^+\to e^+ e^+ \mu^- \nu_e$ vanishes, and so the only contribution comes from the Lepton Number violating process, mediated necessarily by a Majorana $N$. Complementary, the purely
Dirac neutrino $N$ case is shown as the solid line with the label (D).

These figures suggest that measurements of the differential decay width
$d \Gamma/d E^{\prime}_{\mu}$ can more or less distinguish the
Majorana vs. Dirac character of the intermediate neutrino $N$,
the distinction being clearer when the ``admixture parameter''
$\alpha_M$ approaches unity (i.e. $|B_{eN}|^2 \gg |B_{\mu N}|^2$, so that the Lepton Number conserving
component $\pi^+\to e^+ e^+ \mu^- \nu_e$ is relatively suppressed in the Majorana case).
Furthermore, comparison of the results
in figures \ref{dGdEelCMN} with the corresponding results in figures \ref{dGdEelCMM} shows that the difference can be discerned apparently more clearly in the $N$ rest frame than in the pion rest frame, provided the muon energy can be measured with enough precision:
in the $N$ rest frame the curves for the purely Dirac cases are increasing with energy
$E_{\mu}$ all the way to the upper endpoint, where they suddenly drop to zero (the sharp drop is smoothed out by the finite electron mass), while the curves for the Majorana cases reach earlier a maximum and then drop gradually towards the upper endpoint.
In contrast, in the pion rest frame the muon spectra do not show such a clear distinction between the
Dirac and Majorana cases.

%
%
%
%
%
%
%
Now, concerning the experimental challenges, the determination  of the muon
energy in the lab frame needs to have an uncertainty below 1 MeV to achieve the required level of discrimination. 
This requirement is realistic with current detector technology (for example the momentum resolution for a muon in the inner 
detector of ATLAS is near $10^{-4}$ \cite{Aad:2008zzm}, which means that a 1 GeV muon can be measured with a precision 
of a few times  0.1 MeV). 
Now, in the pion rest frame or in the $N$ rest frame, the muon energy,  $E^{\prime}_{\mu}$ and $E_{\mu}$ respectively, 
is obtained after further kinematic analysis.
In general the former may be obtained with more precision, because it can be inferred just from the
measured muon momentum in the lab frame,
provided an accurate energy of the pion beam is known;
instead, for a precise determination of the muon energy
in the neutrino $N$ frame, a precise measurement of
the primary positron momentum is needed as well. 
In any case the required precision for the positron momentum is also realistic for today detectors. 
Probably the most challenging issue here is to have a pion beam with an energy spread, 
$\delta E_{\pi}/E_{\pi}$ (in the lab frame) which is small enough. 
For highly relativistic pions, the corresponding uncertainty of the muon energy in the pion rest frame is
$\delta E^{\prime}_{\mu}/E^{\prime}_{\mu} \approx \delta E_{\pi}/E_{\pi}$.
This uncertainty needs to be $\lesssim 10^{-2}$, so that the energies $E^{\prime}_{\mu}$ can be determined with
an uncertainty below 1 MeV, as required.
%
%
%
%
%
%

In future high intensity beam facilities, such as Project X at Fermilab, charged pion beams
with lab energies $E_{\pi} \sim 2 -15$ GeV
and a luminosity $\sim 10^{22}$ cm$^{-2}$\,s$^{-1}$
will be produced \cite{Geer}. Taking a beam of 1 cm$^2$, we can expect
$\sim 10^{27}$  pions per day. 
With such a large sample of pion decays, the search for the primary process $\pi^+\to e^+ N$ [cf. eq.\ (\ref{piontoeN})] can already improve the current bound
on the mixing element $|B_{e N}|^2 \lesssim 10^{-8}$ \cite{PIENU:2011aa}  by many orders of magnitude.

According to Table \ref{tab1},
for  $|B_{e N}|^2 \sim  10^{-8}$ we get branching ratios
${\rm Br}^{(M)}(\pi^+ \to e^+ e^+ \mu^- \nu) \sim 10^{-12}$ when the
neutrino mass $m_N$ is in the range between $0.120$ and $0.135$ GeV,
and thus, in principle, $10^{17}$ such
events could be detected per year. However, as explained in the previous
section, the probability of the neutrino $N$ to decay
inside the detector, with the detector of length 
$L \sim 10^1$ m, is
\be
P_N  \sim L/(\gamma_{\pi} c \,\tau_N) \sim \frac{10^{-2}}{\gamma_{\pi}} |B_{e N}|^2
\sim (10^{-4} - 10^{-3}) |B_{e N}|^2 \ ,
\label{PN}
\ee
where $\gamma_{\pi} \sim 10^1 - 10^2$ for $E_{\pi} \approx 2 -15$ GeV. This acceptance factor,
for $|B_{e N}|^2 \sim  10^{-8}$, is thus expected to be $\sim 10^{-12} -10^{-11}$,
meaning that, instead of the $\sim 10^{17}$ events just mentioned, about
$10^5 -10^6$ such events can be detected per year.
Now, if $|B_{e N}|^2$ is smaller by a factor $\sim 10$, the number of such detected events
would be lower by a factor $\sim 10^2$, because $Br \propto |B_{e N}|^2$ and
the acceptance $P_N \propto |B_{e N}|^2$, cf. eqs.~ (\ref{diffBr})-(\ref{totalBr})
and (\ref{PN}).
 Keeping in mind the current upper bounds
for $|B_{e N}|^2 \lesssim 10^{-7}$, we see that the search for the
events of the type $\pi^+ \to e^+ e^+ \mu^- \nu$ is promising, due to the
very large expected number of produced charged pions.
If $|B_{e \mu}|^2 \gtrsim 10^{-10}$, the number of such events would be
$\gtrsim 10^2$ per year, which may allow us to distinguish between the
Majorana and Dirac character of neutrino $N$, cf.~figures \ref{dGdEelCMN}
and \ref{dGdEelCMM}.

%
%

\section{Summary and Conclusions}

We have studied the possibility to discover a sterile neutrino $N$ in the mass range between $m_\mu$ and $m_\pi$, and detect its Majorana or Dirac nature, using the charged pion decays $\pi^+\to e^+ e^+ \mu^-\bar\nu_\mu$ and
$\pi^+\to e^+ e^+ \mu^-\bar\nu_e$. The neutrino in question is in the intermediate state, and in order to have sizable experimental signals, it must be on its mass shell (hence the specified mass range). Even so, the rates are extremely small, so they can only be detected in extremely high intensity pion beam experiments. The first process violates Lepton Number by two units, so it can only be produced if the intermediate neutrino $N$ is Majorana. In contrast, the second process only violates Lepton Flavor, while conserving Lepton Number, so it can be produced indistinctly by a Majorana or Dirac neutrino $N$. However, given that the final neutrino flavor ($\bar\nu_\mu$ or $\nu_e$) is not experimentally observed, both processes can contribute to the measured signal, which we refer to as $\pi^+\to e^+ e^+ \mu^-\nu$. Moreover, the rates of both processes could be comparable, so in general the measurement of the branching ratio $Br(\pi^+\to e^+ e^+ \mu^-\nu)$ allows the discovery of the neutrino $N$, but does not distinguish its Majorana or Dirac nature.

While the rates vanish if the neutrino mass $m_N$ is at one of the ends of the range ($m_\mu$, $m_\pi$), for $m_N$ near the middle of this range the branching ratio can be up to $Br(\pi^+\to e^+ e^+ \mu^-\nu)\sim |B_{\ell N}|^2 \times 10^{-4}$. For the current upper bound on the mixing, namely  $|B_{\ell N}|^2 < 10^{-7}$, this means that the branching ratio can be up to $10^{-11}$.

Since a neutrino $N$ of the required mass must be sterile, its lifetime is rather long, so the decay separates in time (space) as $\pi^+\to e^+ N$ followed by $N\to e^+ \mu^- \nu$. Consequently, the first signal for the discovery of such a neutrino in these kinds of experiments comes simply from the detection of the prompt positron with an energy well below the standard mode $\pi^+\to e^+\nu_e$ in the pion rest frame.

While the decay of this standard mode is chirally suppressed due to the small positron mass, the non-standard primary process $\pi^+\to e^+ N$ is only suppressed by the mixing element $B_{e N}$ (actually looking  for low energy bumps in the positron spectrum is how the current bound on this mixing element was obtained). Consequently, the absence of a low energy bump in the primary positron energy spectrum in these high intensity pion beam experiments will allow to improve by many orders of magnitude the existing upper bounds on the  mixing element $|B_{e N}|$ (currently $|B_{e N}|^2 < 10^{-7}$, for $m_N$ in our range of interest).

The Majorana or Dirac nature of the neutrino $N$ cannot be determined from the branching ratios or from the pure detection of the prompt positron. We have studied the possibility to determine such feature from the energy distribution of the muon (a muon with charge opposite to that of the decaying pion). We find that the muon spectrum in the rest frame of the intermediate neutrino $N$ is quite different for the two processes $\pi^+\to e^+ e^+ \mu^-\bar\nu_\mu$ and $\pi^+\to e^+ e^+ \mu^-\bar\nu_e$: while the first reaches a maximum and then decreases gradually to zero at the upper endpoint, the second grows monotonically all the way up to the endpoint, where it sharply drops to zero. If $N$ is a Dirac neutrino, the second feature will clearly show, as the first process will be forbidden (a Dirac neutrino cannot induce Lepton Number violation). On the other hand, if $N$ is a Majorana neutrino, both processes will occur and the distinction could be less clear, depending on the mixing parameters. If $|B_{eN}| \gg |B_{\mu N}|$ (i.e. $\alpha_M\to 1$), the Lepton Number violating process $\pi^+\to e^+ e^+ \mu^-\bar\nu_\mu$ dominates, and the spectrum will show more clearly its shape, signaling the presence of a Majorana neutrino. On the other hand, if $|B_{eN}| \ll |B_{\mu N}|$, it is the Lepton Number conserving process that dominates, so that even if $N$ is Majorana, the spectrum will show the same shape as if $N$ were a Dirac neutrino. In either case, a good energy resolution will be important to tackle the separation between the Dirac and Majorana nature of the neutrino $N$ using the muon spectrum.


%
%

\begin{acknowledgments}
\noindent
We thank Will Brooks for useful discussions. 
The authors acknowledge support from CONICYT (Chile) Ring ACT119 project. 
G.C. and C.S.K. also received support from FONDECYT (Chile) grant
1095196.
The work of C.S.K. was supported in part by the NRF
grant funded by the Korean government of the MEST
(No. 2011-0027275),  (No. 2012-0005690) and (No. 2011 -0020333).

\end{acknowledgments}

%
%

\appendix

\section{Appendix:  general formulas for $M^+ \to \ell_1^+ N \to \ell_1^+ \ell_2^+ \ell^- \nu$ in the $M^+$ and $N$  rest frames}

In this Appendix we present general formulas for the
the Lepton Number violating (LNV) and the Lepton Number conserving (LNC)
decay of a charged pseudoscalar
meson $M^+$, $M^+(p_M) \to \ell_1^+(p_1) \ell_2^+(p_2) \ell^-(p_{\ell}) {\nu}(p_{\nu})$,
cf.~figures \ref{FigPiMaj} and \ref{FigPiDir}, respectively.
Both decays are assumed to take place via the exchange of an
on-shell neutrino $N$ which is assumed to be, in general, Majorana.
The masses of the particles are assumed in general to be nonzero,
exccept for the mass of the standard neutrino ($m_{\nu} = 0$).
For the direct channel of the
LNV decay $M^+ \to \ell^+_1 \ell^+_2 \ell^- {\bar \nu}_{\ell}$
(cf.~figure \ref{FigPiMaj}) the transition amplitude is
\begin{equation}
{\cal M} = \frac{i G_F^2 (B_{\ell_1 N}^{*} B_{\ell_2 N}^{*} 
\lambda_N^{*} ) V_{q Q}^{*} f_M m_N}{p_N^2-m_N^2+i m_N \Gamma_N}\ \
\left[ {\overline u_{\ell}}(p_{\ell}) \gamma^{\eta} 
(1 - \gamma_5) v_{\nu}(p_{\nu}) \right] 
\left[{\overline v^c_{\ell_2}}(p_2) \gamma_{\eta} 
{\displaystyle{\not}p}_M (1 - \gamma_5)
v_{\ell_1}(p_1) \right] \ .
\label{amplitude1}
\end{equation}
We use the notations: 
$\Gamma_N$ is the total decay width of the
neutrino $N$; $G_F$ is the Fermi coupling constant
($G_F \approx 1.166 \cdot 10^{-5} \ {\rm GeV}^{-2}$); 
$B_{\ell_j N}$ is the mixing element between the neutrino of flavor state 
$\nu_{\ell_j}$ with the (mass eigenstate) neutrino $N$, 
cf.~eq.~(\ref{mixing}); $\lambda_N$ is the phase factor of the 
Majorana neutrino $N$ ($|\lambda_N|=1$); $f_M$ is the decay constant
of the meson $M^+$; and
$V_{q Q}$ is the Cabibbo-Kobayashi-Maskawa (CKM) matrix element corresponding 
to $M^{+}$. The subscript $c$ in the spinor for $\ell_2^+$ in 
eq.~(\ref{amplitude1}) means the charge-conjugation 
($v^c = u$ in the helicity basis). 
Furthermore, we will denote by $m_j$ the mass of $\ell^+_j$ lepton ($j=1,2$);

Since the process is dominated by the intermediate neutrino $N$ on 
mass shell, its propagator in the transition probability $|{\cal M}|^2$ 
can be given by the {narrow width approximation}:
\begin{equation}
 \frac{1}{(p_N^2-m_N^2)^2  +  m_N^2 \Gamma_N^2} \simeq
 \frac{\pi}{m_N\Gamma_N} \delta (p_N^2-m_N^2) .
\label{Narrow}
\end{equation}
For the calculation of the decay rate, $\Gamma(M^+\to \ell_1^+ \ell_2^+ \ell^-
\bar\nu_\ell)  =  (2 m_M)^{-1}
\int d_{ps4} |{\cal M}|^2$, it is convenient to decompose the 
4-body phase space integral as the 2-body phase space
of $p_1$ and $p_N$ (where the intermediate momentum $p_N$ has a variable mass 
$\mu_N$), and the 3-body phase space of
$p_2$, $p_\ell$ and $p_\nu$ following from the $N$ decay:
\begin{eqnarray}
 \int d_{ps4} =
 \int \frac{d {\mu_N}^2}{2\pi}
 &&
 \int \frac{d^3 p_1}{(2\pi)^3 2E_{1}} \frac{d^3 p_N}{(2\pi)^3 2E_{N}}
 (2\pi)^4\delta^4(p_M-p_1-p_N) \\
 && \times \int \frac{d^3 p_2}{(2\pi)^3 2E_{2}} 
\frac{d^3 p_\ell}{(2\pi)^3 2E_\ell}
 \frac{d^3 p_\nu}{(2\pi)^3 2E_\nu} (2\pi)^4\delta^4(p_N-p_2-p_\mu-p_\nu) ,
\nonumber
\end{eqnarray}
where the range of the $\mu_N$ integration is  
$\mu_N\in (m_\ell+m_2, m_M - m_1)$. This is precisely the range of possible 
neutrino masses $m_N$ where $N$ can be on mass shell in the process. 
We can now integrate the probability over the phase space. Since the 
integral expression is Lorentz invariant, we can choose any reference frame. 
It turns out that, in the narrow width approximation
(\ref{Narrow}), the direct channel (figure \ref{FigPiMaj})
of the LNV process and the crossed process ($p_1 \leftrightarrow p_2$
and $m_1 \leftrightarrow m_2$) contribute as a sum to $\Gamma$, i.e., the 
interference term is zero. 

\begin{figure}[htb] 
\centering\includegraphics[width=80mm]{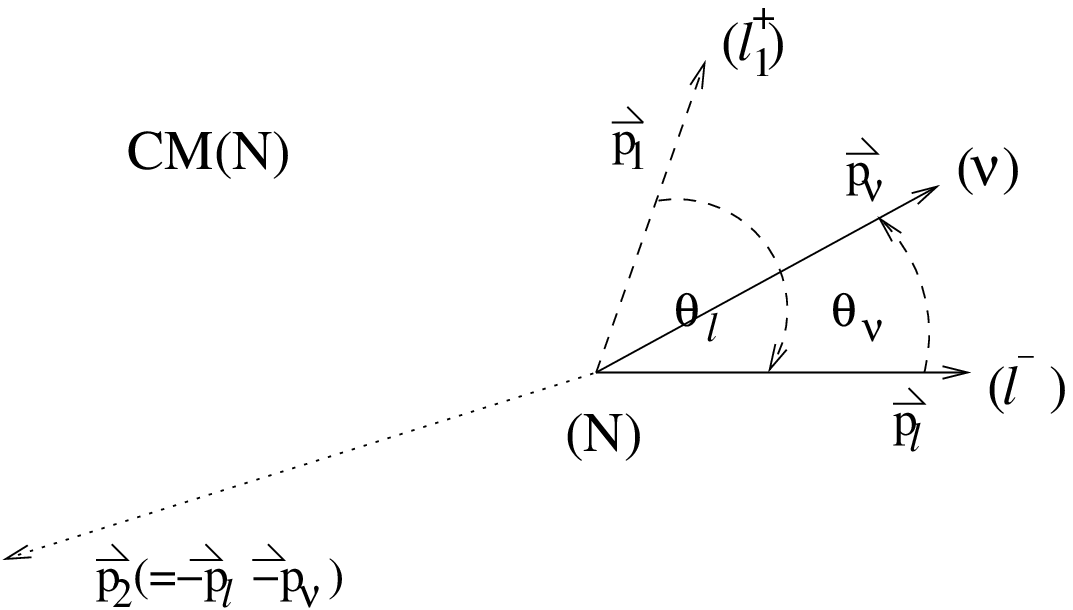}
\caption{\footnotesize}
\label{CMN}
\end{figure}

We work, for convenience, 
in the center-of-momentum (rest) frame of $N$, CM($N$),
where $p_N=(m_N,{\vec 0})$.\footnote{
Nonetheless, the (differential) decay widths $d \Gamma$ will be written for the
$M^+$ rest frame, i.e., without the time dilation factor.}
We choose the ${\hat z}$ direction
along ${\vec p}_{\ell}$, denote the angle between ${\vec p}_{\ell}$ and ${\vec p}_{\nu}$
as $\theta_{\nu}$, and the angle  between ${\vec p}_{\ell}$ and ${\vec p}_1$
as $\theta_{\ell}$, see figure \ref{CMN}
(we refer to the diagram of the type of figure \ref{FigPiMaj}).
For the LNV decay $M^+ \to \ell_1^+ \ell_2^+ \ell^- {\bar \nu}_{\ell}$ 
(cf.~figure \ref{FigPiMaj} and the crossed diagram)
we obtain  (in CM($N$))
\bea
\frac{d \Gamma^{(LNV)}}{d E_{\ell} d \cos \theta_{\ell}} &=&
Z^{(LNV)} \frac{\left[ m_N (m_N - 2 E_{\ell}) + m_{\ell}^2 - m_2^2 \right]^2}{
4 m_N \left[  m_N (m_N- 2 E_{\ell}) + m_{\ell}^2 \right]}
\nonumber\\
&& \times
{\bigg \{}
\cos \theta_{\ell}
(m_N^2 -m_1^2) \left[
 (m_M^2 - m_N^2)^2 - 2 m_1^2 (m_M^2 + m_N^2) + m_1^4 \right]^{1/2}
 (E_{\ell}^2 - m_{\ell}^2)
\nonumber\\
&& +
\left[ m_N^2 (m_M^2 - m_N^2) + m_1^2 (m_M^2 + 2 m_N^2) - m_1^4
\right] E_{\ell} |{\vec p}_{\ell}|
{\bigg \}}
+ (m_1 \leftrightarrow m_2) \ ,
\label{dGM}
\eea
where $Z^{(LNV)}$ is defined as
\be
Z^{(LNV)} \equiv  \left( 1 - \frac{1}{2} \delta_{\ell_1,\ell_2} \right)
 G_F^4 |B_{\ell_1 N}^{*} B_{\ell_2 N}^{*} \lambda_N^{*}  V_{q Q}^{*}|^2
\frac{2}{(2 \pi)^4} \frac{m_N f_M^2}{ \Gamma_N m_M^3}
 \lambda^{1/2}(m_M^2, m_N^2, m_1^2) \ ,
\label{ZM}
\ee
and $\lambda^{1/2}$ is the square root of the function
\be
\lambda(x,y,z) \equiv x^2 + y^2 + z^2 - 2 xy - 2 yz - 2 zx \ .
\label{lamdef}
\ee
We note that the factor $( 1 -  \delta_{\ell_1,\ell_2}/2)$ in $Z^{(LNV)}$
accounts for the fact that, when $\ell_1^+ \not= \ell_2^+$, there are
two types of decays (with intermediate on-shell $N$) leading to the result
$M^+ \to \ell_1^+ \ell_2^+ \ell^- {\bar \nu}_{\ell}$:
(a) $M^+ \to \ell_1^+ N \to \ell_1^+ \ell_2^+ \ell^- {\bar \nu}_{\ell}$, and
(b) $M^+ \to \ell_2^+ N \to \ell_2^+ \ell_1^+ \ell^- {\bar \nu}_{\ell}$.

Integration over the angle $\theta_{\ell}$ gives
\bea
\lefteqn{\frac{d \Gamma^{(LNV)}}{d E_{\ell}}
(M^+ \to \ell_1^+ \ell_2^+ \ell^- {\bar \nu}_{\ell} )
=  Z^{(LNV)} \frac{1}{2 m_N}
\left[ m_M^2 (m_N^2 + m_1^2) - (m_N^2 - m_1^2)^2 \right]}
\nonumber\\
&& \times
E_{\ell} \sqrt{E_{\ell}^2 - m_{\ell}^2}
\frac{\left(m_N^2 - 2 m_N E_{\ell} + m_{\ell}^2 - m_2^2 \right)^2}
{ \left(m_N^2 - 2 m_N E_{\ell} + m_{\ell}^2 \right)}
+ (m_1 \leftrightarrow m_2) \ ,
\label{dGMEel}
\eea
where the first (``d'') channel term
(for: $M^+ \to \ell_1^+ N \to \ell_1^+ \ell_2^+ \ell^- {\bar \nu}_{\ell}$)
contributes when
$m_{\ell} \leq E_{\ell} \leq (E_{\ell})^{\rm (d)}_{\rm max}$,
and the second (``c'') channel term
(for:  $M^+ \to \ell_2^+ N \to \ell_2^+ \ell_1^+ \ell^- {\bar \nu}_{\ell}$)
contributes when
$m_{\ell} \leq E_{\ell} \leq (E_{\ell})^{\rm (c)}_{\rm max}$, where
\bea
(E_{\ell})^{\rm (d)}_{\rm max} &=&
\frac{1}{2 m_N} \left( m_N^2 + m_{\ell}^2 - m_2^2 \right) \ ,
\qquad
(E_{\ell})^{\rm (c)}_{\rm max} =
\frac{1}{2 m_N} \left( m_N^2 + m_{\ell}^2 - m_1^2 \right) \ .
\label{Eelmax}
\eea
When the masses $m_1$ and $m_2$ are negligible, we obtain
\be
\frac{d \Gamma^{(LNV)}}{d E_{\ell} d \cos \theta_{\ell}}{\bigg |}_{m_1=m_2=0}
=Z^{(LNV)}
\frac{1}{2} m_N (m_M^2 - m_N^2) |{\vec p}_{\ell}|
\left[ m_{\ell}^2 + m_N (m_N - 2 E_{\ell}) \right]
\left[ \cos \theta_{\ell} |{\vec p}_{\ell}| + E_{\ell} \right] ,
\label{dGMM0}
\ee
\be
\frac{d \Gamma^{(LNV)}}{d E_{\ell}}{\bigg |}_{m_1=m_2=0}
=Z^{(LNV)}  m_N (m_M^2 - m_N^2)  E_{\ell}  \sqrt{E_{\ell}^2 - m_{\ell}^2}
\left(m_N^2 - 2 m_N E_{\ell} + m_{\ell}^2 \right) \ ,
\label{dGMdEM0}
\ee
\be
\Gamma^{(LNV)}(M^+ \to \ell_1^+ \ell_2^+ \ell^- {\bar \nu}_{\ell} )
{\big |}_{m_1=m_2=0}
=  Z^{(LNV)} \frac{m_N^8}{96}  \left( \frac{m_M^2}{m_N^2} - 1 \right)
f(m_{\ell}^2/m_N^2) \ ,
\label{GMM1M20}
\ee
where $f(x)$ is the 3-body decay function, eq.~(\ref{Ffunction}).

For the LNC decay
$M^+ \to \ell^+_1 \ell^+_2 \ell^- \nu_j$\footnote{the cases $j=1,2$ were added}
(cf.~figure \ref{FigPiDir} and the crossed diagram) we obtain
\bea
\frac{d \Gamma^{(LNC)}}{d E_{\ell} d \cos \theta_{\ell}} &=&
Z^{(LNC)}
\frac{(-1) |{\vec p}_{\ell}|  \left[ -m_2^2 + m_{\ell}^2 +
     m_N (m_N - 2 E_{\ell}) \right]^2}{24 m_N \left[m_{\ell}^2 +
     m_N (m_N - 2 E_{\ell}) \right]^3}
{\bigg \{}
\nonumber\\ &&
\cos \theta_{\ell} (m_1^2 - m_N^2)  |{\vec p}_{\ell}|
\sqrt{((m_M+m_1)^2 - m_N^2) ((m_M-m_1)^2 - m_N^2)}
\nonumber\\ &&
\times
{\big [}
\left( 3 m_{\ell}^2 + m_N (m_N - 4 E_{\ell}) \right)
\left(   m_{\ell}^2 + m_N (m_N - 2 E_{\ell}) \right)
+ m_2^2
\left( 3 m_{\ell}^2 - m_N (m_N + 2 E_{\ell}) \right)
{\big ]}
\nonumber\\ &&
 +
{\big [}
\left( m_1^4 - m_N^2 (m_M^2 - m_N^2) - m_1^2 (m_M^2 + 2 m_N^2) \right)
{\big (}
8 E_{\ell}^3 m_N^2 - 2 m_{\ell}^2 m_N (2 m_2^2 + m_{\ell}^2 + m_N^2)
\nonumber\\ &&
+ 2 E_{\ell}^2 m_N (m_2^2 + 5 m_{\ell}^2 + 5 m_N^2)
 + E_{\ell} ( 3 m_2^2 m_{\ell}^2 + 3 m_2^2 m_N^2 +
(3 m_{\ell}^2 + m_N^2) (m_{\ell}^2 + 3 m_N^2)) {\big )}
{\big ]}
{\bigg \}}
\nonumber\\
&& + (m_1 \leftrightarrow m_2, \ell_1 \leftrightarrow \ell_2) \ ,
\label{dGD}
\eea
where $Z^{(LNC)}$ is defined as
\be
Z^{(LNC)} \equiv  G_F^4
|B_{\ell_1 N}^{*} B_{\ell N}  V_{q Q}^{*}|^2
\left( 1 - \frac{1}{2} \delta_{\ell_1,\ell_2} \right)
\frac{2}{(2 \pi)^4} \frac{ m_N f_M^2}{\Gamma_N m_M^3}
 \lambda^{1/2}(m_M^2, m_N^2, m_1^2) \ .
\label{ZD}
\ee
Integration over $\theta_{\ell}$ gives
\bea
\lefteqn{
\frac{d \Gamma^{(LNC)}}{d E_{\ell}}(M^+ \to \ell_1^+ \ell_2^+ \ell^- {\nu} )
=  Z^{(LNC)} \frac{1}{96 m_N^2}
\frac{1}{\left[ m_{\ell}^2 + m_N (-2 E_{\ell} + m_N) \right]^3}}
\nonumber\\
&&  \times
{\bigg \{} 8 \sqrt{(E_{\ell}^2 - m_{\ell}^2)}
   m_N \left[ m_2^2 - m_{\ell}^2 + (2 E_{\ell} - m_N) m_N \right]^2
\nonumber\\
&& \times
\left[ -m_1^4 + m_M^2 m_N^2 - m_N^4 + m_1^2 (m_M^2 + 2 m_N^2) \right]
{\big [} 8 E_{\ell}^3 m_N^2 - 2 m_{\ell}^2 m_N (2 m_2^2 + m_{\ell}^2 + m_N^2)
\nonumber\\
&&
- 2 E_{\ell}^2 m_N \left( m_2^2 + 5 (m_{\ell}^2 + m_N^2) \right)
+ E_{\ell} (3 m_{\ell}^4 + 10 m_{\ell}^2 m_N^2 + 3 m_N^4 +
3 m_2^2 (m_{\ell}^2 + m_N^2))
{\big ]}
{\bigg \}}
\nonumber\\
&&
+ ( \ell_1 \leftrightarrow \ell_2, m_1 \leftrightarrow m_2 ) \ ,
\label{dGDEel}
\eea
where the explicitly written (``d'') channel term
(for: $M^+ \to \ell_1^+ N \to \ell_1^+ \ell^- \ell_2^+ {\nu}_2$)
has nonzero values
in the interval $m_{\ell} \leq E_{\ell} \leq  (E_{\ell})^{\rm (d)}_{\rm max}$,
and the second (``c'') channel term
(for: $M^+ \to \ell_2^+ N \to \ell_2^+ \ell^- \ell_1^+ {\nu}_1$)
in the interval
$m_{\ell} \leq E_{\ell} \leq  (E_{\ell})^{\rm (c)}_{\rm max}$, where
$(E_{\ell})^{\rm (x)}_{\rm max}$ ($x=d,c$)
are given in eq.~(\ref{Eelmax}).

When the masses $m_1$ and $m_2$ are negligible, we obtain
\bea
\frac{d \Gamma^{(LNC)}}
{d E_{\ell} d \cos \theta_{\ell}}{\bigg |}_{m_1=m_2=0} &=&
\left( Z^{(LNC)} + (\ell_1 \leftrightarrow \ell_2) \right)
\frac{1}{24} m_N (m_M^2-m_N^2) |{\vec p}_{\ell}|
\nonumber\\
&&
\times
{\Big\{}
-3 m_{\ell}^2
\left(
\cos \theta_{\ell} |{\vec p}_{\ell}| + 2 m_N -E_{\ell}
\right)
+ 
\nonumber\\ &&
m_N \left[  \cos \theta_{\ell} |{\vec p}_{\ell}| (4 E_{\ell} - m_N) + (- 4 |{\vec p}_{\ell}|^2 + 3 m_N E_{\ell}) \right]
{\Big\}} \ ,
\label{dGDM0}
\eea
\bea
\frac{d \Gamma^{(LNC)}}{d E_{\ell}}{\bigg |}_{m_1=m_2=0} &=&
\left( Z^{(LNC)} + (\ell_1 \leftrightarrow \ell_2) \right)
\frac{1}{4} m_N (m_M^2-m_N^2) \sqrt{E_{\ell}^2 - M_{\ell}^2}
\nonumber\\
&& \times
\left[ (m_N^2 + m_{\ell}^2) E_{\ell} - \frac{2}{3} m_N (E_{\ell}^2 + m_{\ell}^2) \right] \ ,
\label{dGDdEM0}
\eea
\be
\Gamma^{(LNC)}(M^+ \to \ell_1^+ \ell_2^+ \ell^- {\nu} ) {\big |}_{m_1=m_2=0}
=
\frac{1}{2} \left( Z^{(LNC)} + (\ell_1 \leftrightarrow \ell_2) \right)
 \frac{m_N^8}{96}  \left( \frac{m_M^2}{m_N^2} - 1 \right)
f(m_{\ell}^2/m_N^2) ,
\label{GDM1M20}
\ee
the last expression being almost identical with eq.~(\ref{GMM1M20})
for the LNV decays,
except that now $Z^{(LNV)}$ of eq.~(\ref{ZM}) is replaced by $Z^{(LNC)}$ of
eq.~(\ref{ZD}).

The general case of Majorana ($M$) neutrino $N$ includes LNV and LNC decays, 
and in such a case $d \Gamma^{(M)} = d \Gamma^{(LNV)} + d \Gamma^{(LNC)}$ (if
$\ell_1 \not= \ell$ and $\ell_2 \not=\ell$). 

We note that the case of negligible masses  $m_1$ and $m_2$ 
and with $\ell_1 = \ell_2$
is applicable to the decays considered in the main text of this work,
namely $\ell_1^+ = \ell_2^+ = e^+$; $\ell^- = \mu^-$; and $M^+ = \pi^+$.
Specifically, the expressions
(\ref{dGMdEM0})-(\ref{GMM1M20}) and (\ref{dGDdEM0})-(\ref{GDM1M20})
acquire in this case the following explicit form:
\begin{eqnarray}
\lefteqn{\Gamma(\pi^+\to e^+e^+\mu^-\bar\nu_\mu) = \int_{m_\mu}^{ \frac{m_N^2 + m_\mu ^2 }{2 m_N}  }
  d E_{\mu} }
\label{spectrumLNV}\\
 && \qquad \frac{ G_F^4  |B_{eN}|^4 f_\pi^2 \, |V_{ud}|^2 \, m_N^2\,  (m_\pi^2- m_N^2)^2 }{16\pi^4  \,  m_\pi^3 \,  \Gamma_N}
  \  E_{\mu}\
  (m_N^2 + m_\mu^2- 2 m_N  E_{\mu})
  \sqrt{E_{\mu}^2 - m_\mu^2},
\nonumber
\end{eqnarray}
\begin{equation}
\Gamma(\pi^+\to e^+e^+\mu^-\bar\nu_\mu) =
  \frac{ G_F^4  |B_{eN}|^4 f_\pi^2 \, |V_{ud}|^2 \, m_N^2 \, 
(m_\pi^2- m_N^2)^2 }{16\pi^4  \,  m_\pi^3 \,  \Gamma_N}\cdot
 \frac{m_N^5}{96}\  f\left(m_\mu^2/m_N^2\right) \ ,
\label{maindecay}
 \end{equation}
\begin{eqnarray}
\lefteqn{\Gamma (\pi^+\to e^+\mu^- e^+\nu_e)
= \int_{m_\mu}^{\frac{m_N^2+m_\mu^2}{2 m_N}} dE_{\mu}
}
\label{spectrumLFV}\\
&&\frac{ G_F^4 |B_{eN}B_{\mu N}^\ast|^2 f_\pi^2\, |V_{ud}|^2 \, 
m_N^2(m_\pi^2-m_N^2)^2}
{32\pi^4  \, m_\pi^3  \, \Gamma_N}
 \left\{(m_N^2+m_\mu^2)E_{\mu}
-\frac{2}{3}m_N(2E_{\mu}^2+m_{\mu}^2)\right\}\sqrt{E_{\mu}^2-m_{\mu}^2},
\nonumber
\end{eqnarray}
\begin{equation}
\Gamma (\pi^+\to e^+\mu^- e^+\nu_e)
= \frac{ G_F^4 |B_{eN}B_{\mu N}^\ast|^2 f_\pi^2\, |V_{ud}|^2 \, 
m_N^2(m_\pi^2-m_N^2)^2}
{16\pi^4  \, m_\pi^3  \, \Gamma_N}
\cdot
 \frac{m_N^5}{96}\  f\left(m_\mu^2/m_N^2\right), \label{backdecay}
 \end{equation}
where $f(x)$ is given in eq.\ (\ref{Ffunction}) and $\Gamma_N$ is given in 
eq.~(\ref{DNwidth}).

The differential distributions $d \Gamma/d E_{\ell}$ for the decay
$M^+ \to \ell_1^+ \ell_2^+ \ell^- {\bar \nu}_{\ell} (\nu_j)$,
eqs.~(\ref{dGMEel}) and (\ref{dGDEel}),
refer to the quantities $E_{\ell}$ in the center of momentum (rest) frame
of the intermediate neutrino $N$: $\Sigma \equiv$ CM($N$),
see also figure \ref{CMN}, while the $d \Gamma$ refers to the
differential of the decay width in the center of momentum frame (rest)
of the decaying particle $M^+$: $\Sigma^{'} \equiv$ CM($M^+$).
We define the reference axis ${\hat z}$ in both frames as the direction
of ${\vec p}_1$ of $\ell_1^+$, for the decays of the type of figures \ref{FigPiMaj},
\ref{FigPiDir}.

The initial meson $M^+$ is without spin. Therefore, for the direct channel
(cf.~figures \ref{FigPiMaj} and \ref{FigPiDir}) the distribution
$d \Gamma/(d E^{'}_{\ell} d \theta^{'})$ in the frame $\Sigma^{'} \equiv$ CM($M^+$)
is independent of the angle $\theta^{'}$, where  $\theta^{'}$ is the angle
between a chosen reference axis ${\hat z}$ and the direction of
$\ell_1^{+}$ in CM($M^+$)
\begin{figure}[htb] 
\centering\includegraphics[width=150mm]{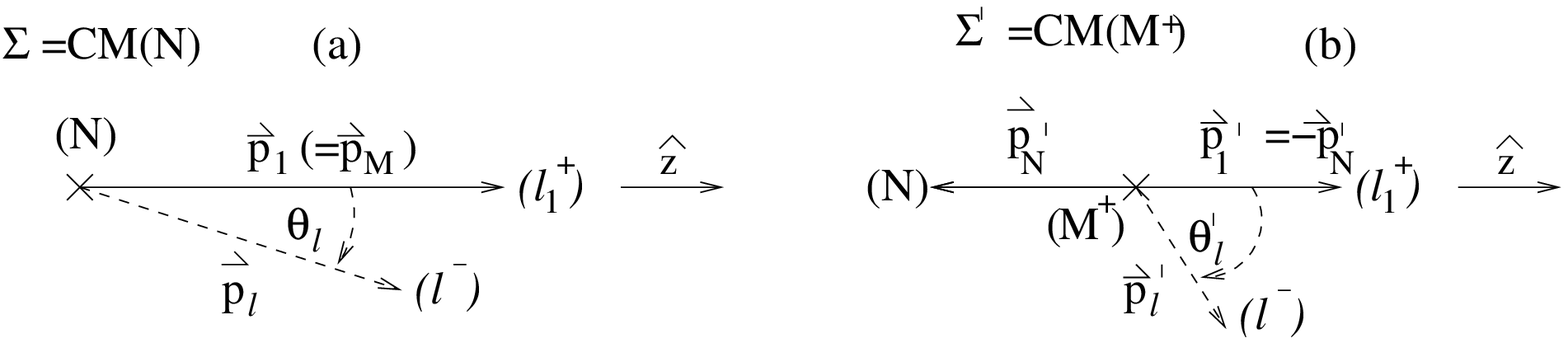}
\caption{\footnotesize}
\label{CMM}
\end{figure}
\be
\frac{d \Gamma}{d E^{'}_{\ell} d \Omega(\theta^{'},\phi^{'})} =
\frac{1}{4 \pi} \frac{d \Gamma}{d E^{'}_{\ell}} \ .
\label{thindep}
\ee
Therefore, we can choose this angle to be
$\theta^{'} = 0$, i.e., ${\vec p^{'}}_1 = |p^{'}_1| {\hat z}$,
cf.~figure \ref{CMM}(b). This axis ${\hat z}$ is the same also in
CM($N$), figure \ref{CMM}(a), i.e., ${\hat z} = {\hat p}_1 = {\hat p}^{'}_1$.

In $\Sigma^{'} \equiv$ CM($M^+$), the momentum $p^{'}_N$ in
the considered direct channel is
\bea
E_N^{'} & = & \frac{1}{2 m_M} ( m_M^2 + m_N^2 - m_1^2) \ ,
\quad
p^{'3}_N =  - |{\vec p}^{'}_N| = - \frac{1}{2 m_M}
\lambda^{1/2}(m_M^2, m_N^2, m_1^2) \ .
\label{pNp}
\eea
The components of $p_{\ell}$ in the two frames $\Sigma \equiv$ CM($N$)
and $\Sigma^{'} \equiv$ CM($M^+$) are then related via a boost determined by
$p^{'}_N$.
The angle between ${\vec p}_{\ell}$ and ${\hat p}_1$ ($={\hat z}$) is
$\theta_{\ell}$ in $\Sigma =$ CM($N$), and
$\theta_{\ell}^{'}$ in $\Sigma^{'} =$ CM($M^+$), i.e.,
$p^3_{\ell} = |{\vec p}_{\ell}| \cos \theta_{\ell}$ and
$p^{'3}_{\ell} = |{\vec p^{'}}_{\ell}| \cos \theta^{'}_{\ell} $.
Then, $E_{\ell}$ and $\cos \theta_{\ell}$ can be expressed as functions
of $E^{'}_{\ell}$, $\cos \theta^{'}_{\ell}$, $E_N^{'}$ and $p^{' 3}_N$, where the
last two quantities are fixed, cf.~eqs.~(\ref{pNp}).
In this way, we obtain the needed one-to-one correspondence
between the energy and the (azimuthal) angle of ${\ell}^-$ lepton
in the two systems CM($N$) and CM($M^+$)
\be
E_{\ell} = E_{\ell}\left( E^{'}_{\ell}, \cos \theta_{\ell}^{'} \right)
=  \frac{1}{m_N} \left(
E^{'}_N E^{'}_{\ell} -
p^{'3}_N \sqrt{ (E_{\ell}^{' 2} - m_{\ell}^2) }  \cos \theta^{'}_{\ell}
\right) \ ,
\label{Eell}
\ee
\bea
\lefteqn{
\cos \theta_{\ell} =
\cos \theta_{\ell} \left(E^{'}_{\ell}, \cos \theta_{\ell}^{'} \right)
}
\nonumber\\
& = &
 \left(  - p^{'3}_N E^{'}_{\ell} +
E^{'}_N  |{\vec p^{'}}_{\ell}| \cos \theta^{'}_{\ell} \right)
\left[
\left(
E^{'}_N E^{'}_{\ell} -
p^{'3}_N \sqrt{(E^{' 2}_{\ell} - m_{\ell}^2)} \cos \theta^{'}_{\ell}
\right)^2 - m_N^2 m_{\ell}^2 \right]^{-1/2} .
\label{costh}
\eea
The distributions $d \Gamma/d E_{\ell}$ and $d \Gamma^{'}/d E_{\ell}^{'}$ are then related
via the following relation:
\be
\frac{d \Gamma}{d E_{\ell}^{'}} =
\int_{c^{'}_{\rm min}}^{c^{'}_{\rm max}}
d \cos \theta_{\ell}^{'} \frac{d^2 \Gamma}{d E_{\ell} d \cos \theta_{\ell}} |J| \ ,
\label{rel1}
\ee
where $c^{'}_{\rm min} \equiv [\cos (\theta_{\ell}^{'}(E_{\ell}^{'}))]_{\rm min}$,
 $c^{'}_{\rm max} \equiv [\cos (\theta_{\ell}^{'}(E_{\ell}^{'}))]_{\rm max}$,
and $J$ is the corresponding Jacobian
\be
J = \frac{\partial ( E_{\ell}, \cos \theta_{\ell} )}
{\partial (E_{\ell}^{'}, \cos \theta_{\ell}^{'})}
= \frac{\partial E_{\ell}}{\partial E_{\ell}^{'}}
\frac{ \partial \cos \theta_{\ell}}{\partial \cos \theta_{\ell}^{'}} -
\frac{\partial E_{\ell}}{\partial \cos \theta_{\ell}^{'}}
\frac{ \partial \cos \theta_{\ell}}{\partial E_{\ell}^{'}} \ ,
\label{Jacob1}
\ee
which can be calculated from the obtained relations (\ref{Eell})-(\ref{costh}).
The result for $J$ is
\bea
J & = & m_N
\left\{
|{\vec p}^{'}_{\ell}| \left(
E_{\ell}^{' 2} E_N^{' 2} - m_{\ell}^2 m_N^2 \right) -
     2 E_{\ell}^{'} E_N^{'} |{\vec p}^{'}_{\ell}|^2 p^{'3}_N \cos \theta_{\ell}^{'} + |{\vec p}^{'}_{\ell}|^3 (E_N^{' 2} - m_N^2) \cos^2 \theta_{\ell}^{'}
\right\}
\nonumber\\
&& \times
\left\{ -m_{\ell}^2 m_N^2 + \left( E_{\ell}^{'} E_N^{'} -
        |{\vec p}^{'}_{\ell}| p^{'3}_N \cos \theta_{\ell}^{'} \right)^2
\right\}^{-3/2}
\label{Jacob2}
\eea
We recall that $E_N^{'}$ and ${\vec p^{' 3}}_N$ are fixed, eqs.~(\ref{pNp}).

The actual differential decay width $d \Gamma/d E_{\ell}^{'}$
as measured in the  $\Sigma^{'}=$ CM($M^+$)
frame is then obtained by integrating the expression (\ref{rel1}) over all the
angles $\theta_{\ell}^{'}$, where $\theta_{\ell}^{'}$ is the angle (in $\Sigma^{'}$)
between $l_1^+$ and ${\ell}^-$ for the decays of the type of
figures \ref{FigPiMaj}, \ref{FigPiDir}, see figure \ref{CMM}(b).
The expression to be integrated over
$\theta_{\ell}^{'}$ is the double differential decay width
$d \Gamma/(d E_{\ell} d \cos \theta_{\ell})$, where
$E_{\ell}$ and $\cos \theta_{\ell}$ are now considered functions of
$E_{\ell}^{'}$ and $\cos \theta_{\ell}^{'}$,
eqs.~(\ref{Eell})-(\ref{costh}).
The double differential decay widths entering the integral
in eq.~(\ref{rel1}) are given in eqs.~(\ref{dGM}) and (\ref{dGMM0})
for the LNV decays, and in  eqs.~(\ref{dGD}) and (\ref{dGDM0})
for the LNC decays.
Since the
relevant parameters in the integration in eq.~(\ref{rel1})
are now $E_{\ell}^{'}$ and $\cos \theta_{\ell}^{'}$,
we also need the relations (\ref{Eell})-(\ref{costh}).

The integration limits in the integral (\ref{rel1}) can be obtained in the
following way. The relations inverse to (\ref{Eell})-(\ref{costh}) are
\bea
E_{\ell}^{'} & = &  \frac{1}{m_N} \left(
E^{'}_N E_{\ell} +
p^{'3}_N |{\vec p}_{\ell}|  \cos \theta_{\ell}
\right)  = E_{\ell}^{'}(E_{\ell}, \cos \theta_{\ell}^{'}) \ ,
\label{Eellp}
\\
\cos \theta_{\ell}^{'} & = & \left(  p^{'3}_N E_{\ell} +
E^{'}_N  |{\vec p}_{\ell}| \cos \theta_{\ell} \right)
\left[
\left(
E^{'}_N E_{\ell} +
p^{'3}_N  |{\vec p}_{\ell}| \cos \theta_{\ell}
\right)^2 - m_N^2 m_{\ell}^2 \right]^{-1/2} \ .
\label{costhp}
\eea
From eqs.~(\ref{Eellp})-(\ref{costhp}) we obtain for
$d \cos \theta_{\ell}^{'}/d E_{\ell}$ at fixed $E^{'}_{\ell}$ a positive
expression, $m_N/ |{\vec p}^{'}_{\ell}|/(-p^{'3}_N)$.
This implies that $E_{\ell}$ grows when $\cos \theta_{\ell}^{'}$ grows, at fixed
$E_{\ell}^{'}$. This, in conjunction with eq.~(\ref{Eell}), means that,
at fixed $E_{\ell}^{'}$, the quantity $m_N E_{\ell}$ varies between
$( E^{'}_N E^{'}_{\ell} + p^{'3}_N |{\vec p}_{\ell}^{'}| )$ and
$( E^{'}_N E^{'}_{\ell} - p^{'3}_N |{\vec p}_{\ell}^{'}| )$.
Simultaneously, we know that $E_{\ell}$ must vary between
$m_{\ell}$ and $(E_{\ell})^{\rm (d)}_{\rm max}$ of eq.~(\ref{Eelmax}) (in the
direct channel). Therefore, $E_{\ell}$, at given fixed $E_{\ell}^{'}$,
varies within and covers the interval which is the
overlap of the two aforementioned intervals, i.e.,
$E_{\ell}(E_{\ell}^{'})_{\rm min} \leq E_{\ell} \leq
E_{\ell}(E_{\ell}^{'})_{\rm max}$, where
\bea
m_N E_{\ell}(E_{\ell}^{'})_{\rm min} &=& {\rm Max} \left(m_N m_{\ell},
E^{'}_N E^{'}_{\ell} + p^{'3}_N |{\vec p}_{\ell}^{'}| \right) \ ,
\label{Eellbound3a}
\\
m_N E_{\ell}(E_{\ell}^{'})_{\rm max} &=& {\rm Min} \left(
\frac{1}{2}(m_N^2 + m_{\ell}^2 - m_2^2),
E^{'}_N E^{'}_{\ell} - p^{'3}_N |{\vec p}_{\ell}^{'}| \right) \ .
\label{Eellbound3b}
\eea
To obtain the corresponding minimal and maximal values of
$\cos \theta_{\ell}^{'}$, at given fixed $E_{\ell}^{'}$, we use an expression
of $\cos \theta_{\ell}^{'}$ in terms of $E_{\ell}^{'}$ and $E_{\ell}$, which is obtained
from eqs.~(\ref{Eellp})-(\ref{costhp})
\be
\cos \theta_{\ell}^{'} =
\frac{1}{\sqrt{ E_{\ell}^{' 2} - m_{\ell}^2}} \frac{1}{(- p_N^{' 3})}
( m_N E_{\ell} - E_N^{'} E_{\ell}^{'} )
= \cos \theta_{\ell}^{'}(E_{\ell}^{'}, E_{\ell}) \ .
\label{cthpEpE}
\ee
Accounting for the aforementioned fact that,
at fixed $E_{\ell}^{'}$, the quantity $\cos \theta_{\ell}^{'}$ grows
when $E_{\ell}$ grows, we finally obtain the following minimal and
maximal values of $\cos \theta_{\ell}^{'}$, at given fixed $E_{\ell}^{'}$,
i.e., the lower and the upper bounds of integration in eq.~(\ref{rel1}):
\bea
c^{'}_{\rm min} & = &
\cos \theta_{\ell}^{'} \left( E_{\ell}^{'}, E_{\ell}(E_{\ell}^{'})_{\rm min} \right)
=
\frac{1}{\sqrt{ E_{\ell}^{' 2} - m_{\ell}^2}} \frac{1}{(- p_N^{' 3})}
\left( m_N E_{\ell}(E_{\ell}^{'})_{\rm min} - E_N^{'} E_{\ell}^{'} \right) ,
\label{cpmin}
\\
c^{'}_{\rm max}
& = & \cos \theta_{\ell}^{'} \left( E_{\ell}^{'}, E_{\ell}(E_{\ell}^{'})_{\rm max} \right)
=  \frac{1}{\sqrt{ E_{\ell}^{' 2} - m_{\ell}^2}} \frac{1}{(- p_N^{' 3})}
\left( m_N E_{\ell}(E_{\ell}^{'})_{\rm max} - E_N^{'} E_{\ell}^{'} \right) ,
\label{cpmax}
\eea
where the energies $E_{\ell}(E_{\ell}^{'})_{\rm min}$ and
$E_{\ell}(E_{\ell}^{'})_{\rm max}$, for a given value of $E_{\ell}^{'}$, are
given in eqs.~(\ref{Eellbound3a})-(\ref{Eellbound3b}). This specifies
the integration limits in the integral (\ref{rel1}).

We note that in the frame $\Sigma^{'}=$ CM($M^+$) the energy of the
$\ell^-$ lepton $E_{\ell}^{'}$ varies between $m_{\ell}$ and
the energy when $E_{\ell}$ is maximal [eq.~(\ref{Eelmax})] and
$\cos \theta_{\ell}=-1$
\be
m_{\ell} \leq E_{\ell}^{'} \leq \frac{1}{2 m_N^2}
\left[ E_N^{'} (m_N^2 + m_{\ell}^2 - m_2^2) +
(- p_N^{' 3}) \lambda^{1/2}(m_N^2,m_{\ell}^2,m_2^2) \right] \ .
\label{Eellpmax}
\ee

Using all these formulas, we can perform explicit numerical evaluations for the
decay $M^+ \to \ell_1^+ \ell_2^+ \ell^{-} \nu_f$
for the case of the pion decay into two positrons:
$M^+ = \pi^+$, $\ell^{-} = \mu^{-}$, $\ell_1^+ = \ell_2^+ = e^+$ ($m_1 = m_2=0$).
Here, $\nu_f = {\bar \nu}_{\mu}, {\nu}_e$ when the intermediate on-shell neutrino
$N$ is Majorana and Dirac, respectively.
The mass $m_N$, for on-shellness, is required to
be between $m_{\mu} = 0.10566$ GeV and $m_{\pi} = 0.13957$ GeV.
Details of the results of such calculations are given in section \ref{num}.
We checked numerically that the integration of the differential decay width
over the energy of the $\mu^{-}$ lepton gave us the same result in the CM($\pi^+$)
as in the CM($N$).
This represents a strong cross-check that our formulas for calculation
of the differential decay width
 $d \Gamma^{\rm (X)}/d E_{\ell}^{'}$ in the meson rest frame CM($M^+$) [=CM($\pi^+$)],
i.e., eq.~(\ref{rel1}) and the subsequent formulas, are correct.

%
%


\begin{thebibliography}{99}

\def\plb#1#2#3{Phys.\ Lett.\       {\bf B#1}, #2 (#3)}
\def\npb#1#2#3{Nucl.\ Phys.\       {\bf B#1}, #2 (#3) }
\def\prd#1#2#3{Phys.\ Rev.\        {\bf D#1}, #2 (#3)}
\def\prl#1#2#3{Phys.\ Rev.\ Lett.\ {\bf #1}, #2 (#3)}
\def\mpl#1#2#3{Mod.\ Phys.\ Lett.\ {\bf A#1}, #2 (#3)}
\def\rep#1#2#3{Phys.\ Rep.\        {\bf #1}, #2 (#3)}
\def\sci#1#2#3{Science             {\bf #1}, #2 (#3)}
\def\astro#1#2#3{Astrophys.\ J.\   {\bf #1}, #2 (#3)}
\def\jcap#1#2#3{JCAP   {\bf #1}, #2 (#3)}
\def\inte#1#2#3{Int.\ J.\ Mod.\ Phy.\   {\bf E#1}, #2 (#3)}

\bibitem{SuperK1}
B. T. Cleveland et al. (Homestake Collaboration), Astrophys. J. {\bf 496}, 505 (1998).

\bibitem{SuperK2}
W. Hampel et al. (GALLEX Collaboration), Phys. Lett. B {\bf 447}, 127 (1999).

\bibitem{SuperK3}
J.N. Abdurashitov et al. (SAGE Collaboration),
Zh. Eksp. Teor. Fiz. {\bf 122}, 211 (2002) [J. Exp. Theor. Phys. 95, {\bf 181} (2002)].

\bibitem{SuperK4}
Y. Fukuda et al. (Super-Kamiokande Collaboration), Phys. Rev. Lett. {\bf 81}, 1562 (1998);
\emph{ibid.} Phys. Rev. Lett. {\bf 86}, 5656 (2001); 
\emph{ibid.}  Phys. Lett. B {\bf 539}, 179 (2002).

\bibitem{SuperK5}
Y. Ashie et al. (Super-Kamiokande Collaboration), Phys. Rev. Lett. {\bf 93}, 101801 (2004).

\bibitem{SuperK6}
Q. R. Ahmad et al. (SNO Collaboration), Phys. Rev. Lett. {\bf 87} 071301 (2001);
\emph{ibid.} Phys. Rev. Lett. {\bf 89}, 011301 (2002). 

\bibitem{SuperK7}
S.N. Ahmed et al. (SNO Collaboration), Phys. Rev. Lett. {\bf 92}, 181301 (2004).

\bibitem{SuperK8}
K. Eguchi et al. (KamLAND Collaboration), Phys. Rev. Lett. {\bf 90}, 021802 (2003).

\bibitem{Racah:1937qq} 
  G.~Racah,
  Nuovo Cim.\  {\bf 14}, 322 (1937).


\bibitem{nndb1}
  W.~H.~Furry,
  Phys.\ Rev.\  {\bf 56}, 1184 (1939).
  
\bibitem{nndb2}  
H.~Primakoff and S.~P.~Rosen, Rep. Prog. Phys. {\bf 22}, 121 (1959);
\emph{ibid.} 
  Phys.\ Rev.\  {\bf 184}, 1925 (1969);
\emph{ibid.} 
  Ann.\ Rev.\ Nucl.\ Part.\ Sci.\  {\bf 31}, 145 (1981).

\bibitem{nndb3}
 M.~Doi, T.~Kotani and E.~Takasugi,
  Prog.\ Theor.\ Phys.\ Suppl.\  {\bf 83}, 1 (1985).
  
\bibitem{Elliot}
  S.~R.~Elliott and J.~Engel,
  J.\ Phys.\ G G {\bf 30}, R183 (2004)
  [hep-ph/0405078].

\bibitem{BB0n1}
V. A. Rodin, A. Faessler, F. Simkovic and P. Vogel, Nucl. Phys. {bf 793}, 213 (2007) .

\bibitem{BB0n2}
M. Kortelainen, O. Civitarese, J. Suhonen and J. Toivanen,  Phys. Lett. B 647 128 (2007).

\bibitem{BB0n3}
M. Kortelainen and J. Suhonen, Phys. Rev. {\bf C 75}, 051303 (2007).

\bibitem{BB0n4}
M. Kortelainen and J. Suhonen, Phys. Rev. {\bf C 76}, 024315 (2007).

\bibitem{BB0n5}
E. Caurier, G. Martinez-Pinedo, F. Nowacki, A. Poves and A. P. Zuker, Rev. Mod. Phys. {\bf 77}, 427 (2005).

\bibitem{BB0n6}
A.~Faessler, G.~L.~Fogli, E.~Lisi, V.~Rodin, A.~M.~Rotunno and F.~Simkovic,
  J.\ Phys.\ G {\bf 35}, 075104 (2008)
  [arXiv:0711.3996 [nucl-th]].

\bibitem{Keung:1983uu} 
  W.~-Y.~Keung and G.~Senjanovi\'c,
  Phys.\ Rev.\ Lett.\  {\bf 50}, 1427 (1983).

\bibitem{Tello:2010am} 
  V.~Tello, M.~Nemev\v{s}ek, F.~Nesti, G.~Senjanovi\'c and F.~Vissani,
  Phys.\ Rev.\ Lett.\  {\bf 106}, 151801 (2011)
  [arXiv:1011.3522 [hep-ph]].

\bibitem{Nemevsek:2011aa} 
  M.~Nemev\v{s}ek, F.~Nesti, G.~Senjanovi\'c and V.~Tello,
  arXiv:1112.3061 [hep-ph].

\bibitem{Senjanovic:2011zz} 
  G.~Senjanovi\'c,
  Riv.\ Nuovo Cim.\  {\bf 034}, 1 (2011).


\bibitem{Littenberg}
  L.~S.~Littenberg and R.~E.~Shrock,
  Phys.\ Rev.\ Lett.\  {\bf 68}, 443 (1992);
 {\it ibid.},
  Phys.\ Lett.\ B {\bf 491}, 285 (2000)
  [hep-ph/0005285].
  
\bibitem{Kova1}
  C.~Dib, V.~Gribanov, S.~Kovalenko and I.~Schmidt,
  Phys.\ Lett.\ B {\bf 493}, 82 (2000)
  [hep-ph/0006277].

\bibitem{Ali}
  A.~Ali, A.~V.~Borisov and N.~B.~Zamorin,
  Eur.\ Phys.\ J.\ C {\bf 21}, 123 (2001)
  [hep-ph/0104123].

\bibitem{deltal2}
  A.~de Gouvea and J.~Jenkins,
  Phys.\ Rev.\ D {\bf 77}, 013008 (2008)
  [arXiv:0708.1344 [hep-ph]].
  
 \bibitem{Atre} 
  A.~Atre, T.~Han, S.~Pascoli and B.~Zhang,
  JHEP {\bf 0905}, 030 (2009)
  [arXiv:0901.3589 [hep-ph]],
and references therein.

\bibitem{cdkk}
  G.~Cveti\v{c}, C.~Dib, S.~K.~Kang and C.~S.~Kim,
  Phys.\ Rev.\ D {\bf 82}, 053010 (2010)
  [arXiv:1005.4282 [hep-ph]].
  
\bibitem{Project-X} {\it Project X and the Science of 
the Intensity Frontier}, white paper based on the 
Project X Physics Workshop, Fermilab, USA, 9-10 November 2009\\
(http://projectx.fnal.gov/pdfs/ProjectXwhitepaperJan.v2.pdf).

\bibitem{Kova}
  J.~C.~Helo, S.~Kovalenko and I.~Schmidt,
  Nucl.\ Phys.\ B {\bf 853}, 80 (2011)
  [arXiv:1005.1607 [hep-ph]].

\bibitem{sdm1}
  A.~D.~Dolgov and S.~H.~Hansen,
  Astropart.\ Phys.\  {\bf 16}, 339 (2002)
  [hep-ph/0009083].

\bibitem{sdm2}
  T.~Asaka, S.~Blanchet and M.~Shaposhnikov,
  Phys.\ Lett.\ B {\bf 631}, 151 (2005)
  [hep-ph/0503065].

\bibitem{sb1}
  A.~Kusenko,
  Phys.\ Rept.\  {\bf 481}, 1 (2009)
  [arXiv:0906.2968 [hep-ph]].

\bibitem{sb2}
  A.~Yu.~Smirnov and R.~Zukanovich Funchal,
  Phys.\ Rev.\ D {\bf 74}, 013001 (2006)
  [hep-ph/0603009].

 \bibitem{cosmology}
  U.~Seljak, A.~Slosar and P.~McDonald,
  JCAP {\bf 0610}, 014 (2006)
  [astro-ph/0604335].

\bibitem{neutrino-osc}
  T.~Schwetz, M.~A.~Tortola and J.~W.~F.~Valle,
  New J.\ Phys.\  {\bf 10}, 113011 (2008)
  [arXiv:0808.2016 [hep-ph]].
  
\bibitem{neutrino-osc2}
  W.~Winter,
  Nucl.\ Phys.\ Proc.\ Suppl.\  {\bf 203-204}, 45 (2010)
  [arXiv:1004.4160 [hep-ph]].

\bibitem{PIENU:2011aa} 
  M.~Aoki {\it et al.}  [PIENU Collaboration],
  Phys.\ Rev.\ D {\bf 84}, 052002 (2011)
  [arXiv:1106.4055 [hep-ex]].

\bibitem{Aad:2008zzm} 
  G.~Aad {\it et al.}  [ATLAS Collaboration],
  JINST {\bf 3}, S08003 (2008).


\bibitem{Geer}
S.~Geer, private communication.
  



\end{thebibliography}
\end{document}